\documentclass[preprint,12pt]{elsarticle}
\usepackage{amssymb,amsmath,graphicx,bm,mathrsfs,color,xcolor,aas_macros,soul}
\biboptions{sort&compress}  
\usepackage{xcolor}
\usepackage[margin=2.5cm]{geometry}  
\newcommand{\be}{\begin{equation}} 
\newcommand{\ee}{\end{equation}}
\newcommand{\bea}{\begin{eqnarray}}
\newcommand{\eea}{\end{eqnarray}}
\newcommand{\Tr}{{\rm Tr}}
\newcommand{\ie}{{\it i.e.}}
\newcommand{\eg}{{\it e.g.}}

\definecolor{red}{rgb}{0.8,0,0}
\definecolor{violet}{rgb}{0.4,0,0.4}
\definecolor{green}{rgb}{0,0.5,0.0}
\definecolor{navy}{rgb}{0.0,0.0,0.6}
\definecolor{orange}{rgb}{0.8,0.2,0.0}

\usepackage[normalem]{ulem}  

\usepackage[colorlinks=true,linkcolor=red,citecolor=red,urlcolor=red]{hyperref}

\journal{Annals of Physics}

\begin{document}
\begin{frontmatter}

\title{{\bf Generalized relativistic second-order dissipative hydrodynamics: 
 coupling different rank tensors}}

\author[label1,label2]{Arus Harutyunyan}
\author[label3,label4]{Armen Sedrakian}
\address[label1]{Byurakan Astrophysical Observatory, Byurakan 0213, Armenia}
\address[label2]{Institute of Physics, Yerevan State University,
  Yerevan 0025, Armenia}
\address[label3]{Institute of Theoretical Physics, University of
  Wroc\l{}aw, 50-204 Wroc\l{}aw, Poland}
\address[label4]{Frankfurt Institute for Advanced Studies, 
  Ruth-Moufang-Str.\ 1, D-60438 Frankfurt am Main, Germany}

\begin{abstract}
 In this work, we extend the formalism of second-order relativistic
  dissipative hydrodynamics, developed previously using Zubarev's
  non-equilibrium statistical operator
  formalism~\cite{Harutyunyan_hydro2022}. By employing a second-order
  expansion of the statistical operator in terms of hydrodynamic
  gradients, we demonstrate that new second-order terms emerge due to
  the coupling of two-point quantum correlators between tensors of
  differing ranks, evaluated at distinct space-time points. Such
    terms arise because of the presence of the acceleration vector in the
    system allows Curie's theorem, which governs symmetry constraints,
    to be extended for constructing invariants from tensors of
    different ranks evaluated at distinct space-time points. The new
  terms are identified in the context of a complete set of
  second-order equations governing the shear stress tensor,
  bulk viscous pressure, and charge diffusion currents for a generic
  quantum system characterized by the energy-momentum tensor and
  multiple conserved charges. Additionally, we identify the transport
  coefficients associated with these new terms and derive the Kubo
  formulas expressing the second-order transport coefficients through
  two- and three-point correlation functions.
\end{abstract}

\end{frontmatter}

\section{Introduction}

\label{sec:hydro_intro}

Hydrodynamics is an effective theory that describes many particle
systems within low-frequency and long-wavelength limits. It finds
numerous applications in astrophysics, nuclear physics, high-energy
physics, etc. Relativistic hydrodynamics has successfully modeled the
collective dynamics of strongly interacting matter generated in
high-energy heavy-ion collision experiments conducted at the
Relativistic Heavy Ion Collider and the Large Hadron
Collider. Additionally, it plays a significant role in the physics of
compact stars, particularly in the study of binary neutron-star
mergers and supernovas.

Relativistic hydrodynamics characterizes the state of a fluid through
its energy-momentum tensor and conserved charge currents. In the
relevant low-frequency and long-wavelength limits, these quantities can
be expanded around their equilibrium values.  To overcome the
acausality and instability in numerical computations found in
first-order theory, second-order relativistic theories were developed
in the late 1970~\cite{1976AnPhy.100..310I,1979AnPhy.118..341I}.  In
these theories, the dissipative fluxes satisfy relaxation equations,
which describe the process of their relaxation towards their
Navier--Stokes values at asymptotically large times.

There are two primary approaches for deriving the equations of
hydrodynamics from the underlying microscopic theory. For weakly
coupled systems, the Boltzmann kinetic theory can be used to determine
the quasi-particle distribution function outside the thermal
equilibrium. In contrast, for strongly interacting systems, a
comprehensive quantum-statistical approach based on the Liouville
equation for the non-equilibrium statistical operator is necessary -
an approach that we will follow below.  Within the class of such
theories, Zubarev's formalism, also known as the method of the
non-equilibrium statistical operator
(NESO)~\cite{zubarev1974nonequilibrium, zubarev1997statistical} allows
to obtain the hydrodynamics equations of a strongly correlated
systems.

This framework expands the traditional Gibbs ensemble methodology to
describe systems beyond equilibrium states. By developing a
statistical operator that captures non-local variations of
thermodynamic parameters and their spatial gradients, one is able to
model system behavior out of equilibrium provided that the
thermodynamic properties change smoothly across characteristic
microscopic correlation lengths. Formally, this amounts to promoting
the statistical operator to a non-local functional of the
thermodynamic parameters and their gradients.  Then, the hydrodynamics
equations for the dissipative currents emerge after full statistical
averaging of the relevant quantum operators.  The NESO method has
garnered significant interest in recent years and has been applied to
relativistic quantum fields and hydrodynamics in
Refs.~\cite{Buzzegoli:2017cqy,Harutyunyan:2018cmm,Tokarchuk:2018jjk,Becattini:2019dxo,Becattini:2020qol,
  Prokhorov:2019sss,Prokhorov:2019cik,Torrieri:2020ezm,Harutyunyan_hydro2022,Granese:2022igc,She2025}.

Second-order relativistic dissipative hydrodynamics was previously
derived using the NESO approach in Ref.~\cite{Harutyunyan_hydro2022},
incorporating terms up to the second order in the expansion of the
statistical operator. This work surpasses earlier studies, which were
constrained to first-order gradient approximations.  This expansion
was shown to be equivalent to an expansion to the second order in the
Knudsen number with the second-order non-local in space-time terms in
the equations governing dissipative currents resulting in nonzero
relaxation time scales. In this paper, we extend the approach
developed in Ref.~\cite{Harutyunyan_hydro2022} to identify additional
non-local contributions.  Such contributions would vanish on naive
application of Curie's theorem, as they couple tensors of
different ranks. However, as we show, the invariance of resulting
correlation functions with respect to symmetry transformations can be
maintained due to the accelerated motion of the fluid.

Before proceeding we note that extensive literature exists on the
derivation of second-order dissipative relativistic hydrodynamics
which utilize alternative expansions (among other attributes), for a
recent review and references
see~\cite{Rocha:2023ilf,Harutyunyan:2023nvt}. Furthermore, recent work
has demonstrated that the observed acausalities and instabilities stem
from the matching procedure to the local-equilibrium reference
state. By generalizing this matching approach, several authors have
derived causal and stable first-order dissipative hydrodynamic
theories~\cite{Bemfica,Kovtun2019}.

The paper is constructed as follows. Section~\ref{sec:Zub_formalism}
provides an overview of the relativistic dissipative hydrodynamics and
Zubarev's formalism for the NESO~\cite{zubarev1974nonequilibrium,
  zubarev1997statistical}. In Sec.~\ref{sec:diss1} we derive the
complete second-order equations, including new terms, for the shear
stress tensor, the bulk viscous pressure, and the diffusion
currents. Our results are summarized and discussed in
Sec.~\ref{sec:hydro_discuss}.  \ref{sec:Relevant_NSO} discusses
the choice of relevant statistical operator in hydrodynamics. Some details related to the
decomposition of the thermodynamic force into different dissipative
processes are provided in \ref{app:C_decomp}. The Kubo formulas and
the relevant correlation functions are derived in
\ref{app:Green_func}.  We work in Minkowski space-time with the
metric $g^{\mu\nu}={\rm diag}(+,-,-,-)$.

\section{The non-equilibrium statistical operator formalism}
\label{sec:Zub_formalism}

In this section, we provide a brief overview of Zubarev's formalism
for the non-equilibrium statistical operator in a quantum system with
multiple conserved charges within the hydrodynamic
regime~\cite{zubarev1997statistical,
  1979TMP....40..821Z,zubarev1974nonequilibrium}. The starting point
of this approach is the conservation laws for the energy-momentum
tensor and the conserved charge currents
\be\label{eq:cons_laws}
\partial_{\mu} \hat{T}^{\mu\nu} =0,\qquad
\partial_{\mu} \hat{N}^{\mu}_a =0,
\ee
where $a=1,2,\dots, \ell$ labels the conserved charges (\eg, 
baryonic, electric, etc.) with $\ell$ being the total number of 
these charges. The equations of relativistic hydrodynamics are 
obtained by averaging these equations over the full non-equilibrium 
statistical operator, which for a multicomponent system is 
given by~\cite{Harutyunyan_hydro2022}
\be\label{eq:stat_op_full}
\hat{\rho}(t) = Q^{-1}e^{-\hat{A}+\hat{B}},
\qquad Q=\Tr e^{-\hat{A}+\hat{B}},
\ee
where the operators $\hat{A}$ and $\hat{B}$ are given by
\bea\label{eq:A_op}
\hat{A}(t)&=&\int\! d^3x \Big[\beta^\nu(x)
\hat{T}_{0\nu}(x)-
\sum_a\alpha_a(x) \hat{N}^0_a(x)\Big],\\
\label{eq:B_op}
\hat{B}(t)&=& 
\int\! d^4x_1\, \hat{C}(x_1),\\
\label{eq:C_op}
\hat{C}(x)&=&\hat{T}_{\mu\nu}(x)
\partial^{\mu}\beta^\nu(x)-
\sum_a\hat{N}^\mu_a(x)\partial_\mu\alpha_a(x),
\eea
where $x\equiv (\bm  x, t)$ denotes a point 
in the space-time,
\bea\label{eq:int_short}
\int d^4x_1 \equiv \int d^3x_1\int_{-\infty}^tdt_1 
e^{\varepsilon(t_1-t)},
\eea
and
\be\label{eq:beta_nu_alpha}
\beta^\nu(x) = \beta(x) u^{\nu}(x),\qquad
\alpha_a (x)=\beta(x)\mu_a (x).
\ee
Here $\beta^{-1}(x)$, $\mu_a(x)$, and $u^{\nu}(x)$ are the local
temperature, the chemical potentials, and the fluid 4-velocity,
respectively. The quantities must be slowly varying functions in space
and time.  This observation applies when macroscopic spatial and
temporal variations of system quantities significantly exceed the
characteristic microscopic scales, such as the mean free path
$\lambda$ of quasi-particles in weakly interacting systems which
  characterizes the range of the interaction. Since the thermodynamic
  parameters and the fluid velocity vary over a {\it macroscopic}
  length scale $L \gg \lambda$, the small parameter with respect to
  which the various orders of expansion are organized is the Knudsen
  number Kn$= \lambda/L \ll 1$, \ie, the first- and second-order
  dissipative hydrodynamics takes into account terms of linear and
  quadratic order in Kn. As explained in \ref{sec:Relevant_NSO}
  the choice of the {\it relevant statistical operator}  implicitly assumes molecular
  chaos which justifies the factorization of the two-particle
  distribution function into a product of single-particle
  distributions for large distances, also used in the kinetic theory to truncate the
  Bogoliubov-Born-Green-Kirkwood-Yvon hierarchy.

In Eq.~\eqref{eq:int_short} one should take the limit
$\varepsilon\to +0$ after the thermodynamic limit is taken.  The
statistical operator given above satisfies the quantum Liouville
equation. An infinitesimal source term is introduced to ensure that
the retarded solutions of the Liouville equation are
chosen~\cite{zubarev1974nonequilibrium,
  zubarev1997statistical,Harutyunyan_hydro2022}.  The operators
$\hat{A}(t)$ and $\hat{B}(t)$ represent the equilibrium and
non-equilibrium components of the statistical operator,
respectively. The operator $\hat{C}(x)$ serves as the thermodynamic
 ``force'' operator, aggregating gradients of key thermodynamic variables
including temperature, chemical potentials, and fluid 4-velocity.

In the next step, we expand the statistical operator in 
power series with respect to the thermodynamic force 
$\hat{B}(t)$ up to the second order~\cite{Harutyunyan_hydro2022}
\bea  \label{eq:stat_full_2nd_order}
\hat{\rho} = \hat{\rho}_l+\hat{\rho}_1+\hat{\rho}_2,
\eea
where $\hat{\rho}_l=e^{-\hat{A}}/\Tr e^{-\hat{A}}$ is the local
equilibrium part of the statistical operator, also referred to as
relevant statistical operator~\cite{zubarev1974nonequilibrium,
  zubarev1997statistical}. As seen from Eq.~\eqref{eq:A_op},
$\hat{\rho}_l$ is the generalization of the Gibbs distribution for
local equilibrium states.

The first- and the second-order corrections are given, respectively, by
\bea\label{eq:rho_1_final}
\hat{\rho}_1 (t) =  \int d^4x_1\int_0^1d\lambda 
\left[\hat{C}_\lambda(x_1) - \langle \hat{C}_\lambda(x_1)
\rangle_l\right]\hat{\rho}_{l},
\eea
and 
\bea\label{eq:rho_2_final}
\hat{\rho}_2 (t) 
= \frac{1}{2}\int\! d^4x_1d^4x_2\!
\int_0^1\!  d\lambda_1\! \int_0^1\! d\lambda_2
 \Big[\tilde{T} \{\hat{C}_{\lambda_1}(x_1)
\hat{C}_{\lambda_2}(x_2)\} -\big\langle 
\tilde{T}\{ \hat{C}_{\lambda_1}(x_1)
\hat{C}_{\lambda_2}(x_2)\}\big\rangle_l \nonumber\\
- \big\langle \hat{C}_{\lambda_1}(x_1)\big\rangle_l 
 \hat{C}_{\lambda_2}(x_2) -\hat{C}_{\lambda_1}(x_1)
\big\langle \hat{C}_{\lambda_2}( x_2)\big\rangle_l  
+ 2 \big\langle \hat{C}_{\lambda_1}( x_1)\big\rangle_l 
\big\langle \hat{C}_{\lambda_2}(x_2)\big\rangle_l\Big]\hat{\rho}_{l},
\eea
where we defined
$\hat{X}_\lambda = e^{-\lambda A} \hat{X}e^{\lambda A}$ for any
operator $\hat{X}$, and $\tilde{T}$ is the anti-chronological operator
for $\lambda$ variables.  The statistical average of any operator
$\hat{X}(x)$ can be now written according to
Eqs.~\eqref{eq:stat_full_2nd_order}, \eqref{eq:rho_1_final} and
\eqref{eq:rho_2_final} as
\bea\label{eq:stat_average}
\langle \hat{X}(x)\rangle =\Tr[\hat{\rho}(t)\hat{X}(x)]
&=& \langle \hat{X}(x)\rangle_l +
\int\! d^4x_1
\Big(\hat{X}(x),\hat{C}(x_1)\Big) \nonumber\\
&+& \int\! d^4x_1d^4x_2
\Big(\hat{X}(x),\hat{C}(x_1),\hat{C}(x_2)\Big)+\dots,
\eea
where $\langle \hat{X}(x)\rangle_l =\Tr[\hat{\rho}_l(t)\hat{X}(x)]$ is
the local-equilibrium average, and we defined the two-point
correlation function by
\bea\label{eq:2_point_corr} \Big(\hat{X}(x),\hat{Y}(x_1)\Big) \equiv
\int_0^1\! d\lambda\, \Big\langle\hat{X}(x) \left[\hat{Y}_\lambda(x_1)
  - \big\langle \hat{Y}_\lambda(x_1)\big\rangle_l\right]\Big\rangle_l,
\eea
and the three-point correlation function by
\bea\label{eq:3_point_corr}
\Big(\hat{X}(x),\hat{Y}(x_1),\hat{Z}(x_2)\Big) &\equiv&
\frac{1}{2} \int_0^1\!  d\lambda_1\! \int_0^1\! d\lambda_2 \, \Big\langle \tilde{T} \left\{
\hat{X}(x)\Big[\hat{Y}_{\lambda_1}(x_1)\hat{Z}_{\lambda_2}(x_2) \right. \nonumber\\
&-&
 \big\langle \hat{Y}_{\lambda_1}(x_1)\big\rangle_l  
\hat{Z}_{\lambda_2}(x_2)- \hat{Y}_{\lambda_1}( x_1) 
\big\langle\hat{Z}_{\lambda_2}(x_2)\big\rangle_l \nonumber\\
&-& \left.
\big\langle\tilde{T}\,\hat{Y}_{\lambda_1}(x_1)\hat{Z}_{\lambda_2}(x_2)\big\rangle_l+
2\big\langle \hat{Y}_{\lambda_1}( x_1)\big\rangle_l
\big\langle \hat{Z}_{\lambda_2}(x_2)\big\rangle_l\Big]\right\}\Big\rangle_l.
\eea
From Eq.~\eqref{eq:3_point_corr}, we derive the following symmetry relation:
\bea\label{eq:3_point_corr_sym}
\Big(\hat{X}(x),\hat{Y}(x_1),\hat{Z}(x_2)\Big) =
\Big(\hat{X}(x),\hat{Z}(x_2),\hat{Y}(x_1)\Big),
\eea
which will be employed in the subsequent analysis.

The following remark is necessary. Thermodynamic variables are
well-defined only in equilibrium states, not in non-equilibrium
ones. However, it is possible to extend the application of these
variables to non-equilibrium states by constructing a hypothetical
local-equilibrium state from which the actual non-equilibrium state
does not depart substantially.  To achieve this, we define the
operators for energy and charge densities in the comoving frame as
$\hat{\epsilon} = u_\mu u_\nu \hat{T}^{\mu\nu}$ and
$\hat{n}_a = u_\mu \hat{N}^\mu_a$.  The local values of the
Lorentz-invariant thermodynamic parameters $\beta$ and $\alpha_a$ are
determined by matching the given average values of the operators
$\hat{\epsilon}$ and $\hat{n}_a$ using the following matching
conditions~\cite{1958PhRv..112.1829M,zubarev1974nonequilibrium,
  1979TMP....40..821Z,
  1987NuPhB.280..716H,zubarev1997statistical,2011AnPhy.326.3075H}
\bea\label{eq:matching}
\langle \hat{\epsilon}(x)\rangle =
\langle\hat{\epsilon}(x)\rangle_l,\qquad
\langle\hat{n}_a(x)\rangle
=\langle\hat{n}_a(x)\rangle_l.
\eea
Note that the temperature and the chemical potentials are defined
according to~\eqref{eq:matching} actually as {\it non-local
  functionals} of $\langle \hat{\epsilon}(x)\rangle\equiv\epsilon(x)$
and
$\langle\hat{n}_a(x)\rangle\equiv n_a(x)$~\cite{1972Phy....59..285Z}.

The thermodynamic parameters can be defined as {\it local} functions
of the energy and charge densities.  For that purpose, the local
equilibrium values $\langle\hat{\epsilon}\rangle_l$ and
$\langle\hat{n}_a\rangle_l$ in Eq.~\eqref{eq:matching} should be
evaluated formally at {\it constant values} of $\beta$ and $\mu_a$.
Then, these can be found by equating
$\langle\hat{\epsilon}\rangle_l$ and $\langle \hat{n}_a\rangle_l$ to
the real values of these quantities $\langle\hat{\epsilon}\rangle$ and
$\langle \hat{n}_a\rangle$ at any particular point in space.  This is
the well-established procedure that allows one to construct a
so-called fictitious local equilibrium state at any given point. It
ensures that one accurately reproduces the local values of energy and
charge densities. Furthermore, this approach also determines the local
values of energy flow or one of the charge currents, depending on
whether Landau's or Eckart's definition of fluid velocity is used, see
Sec.~\ref{sec:eq_hydro}.

\subsection{Equations of relativistic hydrodynamics}
\label{sec:eq_hydro}

To obtain hydrodynamics equations we decompose the energy-momentum
tensor and the charge currents into their equilibrium and dissipative
parts in the standard way by
\bea \label{eq:T_munu_decomp}
\hat{T}^{\mu\nu} &=& \hat{\epsilon} u^{\mu}u^{\nu} - 
\hat{p}\Delta^{\mu\nu} + \hat{q}^{\mu}u^{\nu}+ 
\hat{q}^{\nu}u^{\mu} + \hat{\pi}^{\mu\nu},\\
\label{eq:N_a_decomp}
\hat{N}^{\mu}_a &=& \hat{n}_au^\mu +\hat{j}^{\mu}_a,
\eea
where $\Delta^{\mu\nu}=g^{\mu\nu}-u^\mu u^\nu$ is the projector onto
the 3-space orthogonal to $u_\mu$, the shear stress tensor
$\hat{\pi}^{\mu\nu}$, the energy flux $\hat{q}^\mu$ and the diffusion
currents $\hat{j}^\mu_a$ are orthogonal to $u_\mu$, and
$\hat{\pi}^{\mu\nu}$ is traceless:
\bea \label{eq:orthogonality}
u_{\nu}\hat{q}^{\nu} = 0,\qquad
u_{\nu}\hat{j}^{\nu}_a = 0,\qquad 
u_{\nu}\hat{\pi}^{\mu\nu} = 0,\qquad 
\hat{\pi}_{\mu}^\mu=0.
\eea
Note that the equilibrium and non-equilibrium parts of the pressure
are not separated in Eq.~\eqref{eq:T_munu_decomp}.  The statistical
average of the operator $\hat{p}$ yields the actual thermodynamic
pressure. In nonequilibrium states, this pressure differs from the
equilibrium pressure $p=p\left(\epsilon, n_a\right)$, which is
determined by the equation of state and can be obtained by averaging
the operator $\hat{p}$ over the local equilibrium distribution,
evaluated at constant values of the thermodynamic parameters. The
difference between these two averages represents the non-equilibrium
component of the pressure, known as the bulk viscous pressure, see
Sec.~\ref{sec:diss1} for details.

The operators on the right-hand sides of Eqs.~\eqref{eq:T_munu_decomp}
and \eqref{eq:N_a_decomp} are given by the relevant projections of
$\hat{T}^{\mu\nu}$ and $\hat{N}_a^\mu$
\bea\label{eq:proj1_op}
\hat{\epsilon} = u_\mu u_\nu \hat{T}^{\mu\nu},\qquad
\hat{n}_a = u_\mu\hat{N}^{\mu}_a,\qquad
\hat{p}=-\frac{1}{3}\Delta_{\mu\nu}
\hat{T}^{\mu\nu},\\
\label{eq:proj2_op}
\hat{\pi}^{\mu\nu} = \Delta_{\alpha\beta}^{\mu\nu} 
\hat{T}^{\alpha\beta},\qquad \hat{q}^\mu  = u_\alpha
\Delta_{\beta}^{\mu}\hat{T}^{\alpha\beta},\qquad
\hat{j}_a^{\nu}=\Delta_{\mu}^{\nu} \hat{N}^{\mu}_a,
\eea
where we used the identities
\bea\label{eq:prop_proj}
u_\mu\Delta^{\mu\nu}=\Delta^{\mu\nu}u_\nu=0,\qquad
\Delta^{\mu\nu}\Delta_{\nu\lambda}=\Delta^{\mu}_\lambda,
\qquad \Delta^{\mu}_{\mu}=3,
\eea
and introduced fourth rank traceless 
projector orthogonal to $u^\mu$ via
\bea\label{eq:projector_delta4}
\Delta_{\mu\nu\rho\sigma}= \frac{1}{2}
\left(\Delta_{\mu\rho}\Delta_{\nu\sigma}
+\Delta_{\mu\sigma}\Delta_{\nu\rho}\right)
-\frac{1}{3}\Delta_{\mu\nu}\Delta_{\rho\sigma}.
\eea
Eqs.~\eqref{eq:proj1_op} and \eqref{eq:proj2_op}
 in the local rest frame read 
\bea \label{eq:currents_rest1_op}
\hat{\epsilon} = \hat{T}^{00},\qquad
\hat{n}_a = \hat{N}^{0}_a,\qquad
\hat{p} =-\frac{1}{3}\hat{T}^k_k,\hspace{1.5cm}\\
\label{eq:currents_rest2_op}
\hat{\pi}_{kl} = \left(\delta_{ki}\delta_{lj}-
\frac{1}{3}\delta_{kl}\delta_{ij}
\right) \hat{T}_{ij},\qquad
\hat{q}^i  = \hat{T}^{0i},\qquad
\hat{j}_a^{i} = \hat{N}^{i}_a.
\eea

Substituting Eqs.~\eqref{eq:T_munu_decomp} and \eqref{eq:N_a_decomp}
in Eq.~\eqref{eq:cons_laws} and averaging over the non-equilibrium
statistical operator, we obtain the equations of relativistic
dissipative hydrodynamics
\bea \label{eq:hydro1}
Dn_a +n_a\theta +\partial_\mu j^{\mu}_a &=& 0,\\
\label{eq:hydro2}
D\epsilon + (h+\Pi)\theta +\partial_\mu q^{\mu}-
q^{\mu}D u_\mu -\pi^{\mu\nu}\sigma_{\mu\nu} &=& 0,\\
\label{eq:hydro3}
(h+\Pi) D u_{\alpha}- \nabla_\alpha (p+\Pi)
+q_{\alpha}\theta\nonumber\\
+ q^{\mu}\partial_\mu u_{\alpha} + \Delta_{\alpha\mu}D q^{\mu} 
+\Delta_{\alpha\nu}\partial_\mu \pi^{\mu\nu} &=&0,
\eea
where $\epsilon \equiv \langle \hat{\epsilon} \rangle$,
$n_a \equiv \langle \hat{n}_a \rangle$,
$p+\Pi \equiv \langle \hat{p} \rangle$,
$\pi^{\mu\nu}\equiv \langle \hat{\pi}^{\mu \nu} \rangle$,
$q^\mu \equiv \langle \hat{q}^\mu \rangle$, and
$j_a^{\mu}\equiv\langle \hat{j}_a^\mu \rangle$ are the statistical
averages of the corresponding operators; $p\equiv p(\epsilon,n_a)$
represents the pressure in local equilibrium, \ie, the pressure
determined by the equation of state (EoS), while $\Pi$ denotes the
non-equilibrium component of the pressure (for further details, see
Sec.~\ref{sec:diss1}). The quantity
$h=\epsilon+p=T s+\sum_a \mu_a n_a$ is the enthalpy density, where $T$
is the temperature, $s$ is the entropy density, $\mu_a$ are the
chemical potentials, and $n_a$ are the charge densities. The operator
$D \equiv u^\mu \partial_\mu$ represents the comoving derivative,
which corresponds to the time derivative in the local rest frame,
$\nabla_\alpha\equiv \Delta_{\alpha\beta}\partial^\beta$ is the
covariant spatial derivative,
$\sigma_{\mu\nu}\equiv \Delta_{\mu\nu}^{\alpha\beta} \partial_\alpha
u_\beta$ is the shear stress tensor, and $\theta \equiv \partial_\mu u^\mu$
is the fluid expansion rate.  Equations~\eqref{eq:hydro2} and
\eqref{eq:hydro3} can be obtained by contracting the first
equation of~\eqref{eq:cons_laws} by $u_\nu$ and $\Delta_{\nu\alpha}$,
respectively. The system of Eqs.~\eqref{eq:hydro1}--\eqref{eq:hydro3}
contains $\ell+4$ equations for $4\ell+10$ independent variables. To
close this set of equations additional expressions for the dissipative
fluxes are required.  These consist of $3 \ell$ equations for the
independent components of the diffusion currents, one equation for the
bulk viscous pressure, and 5 equations for the independent components
of the shear stress tensor.  We remind here that the energy flow or
one of the diffusion currents can be eliminated from hydrodynamics
equations by a relevant choice of the fluid 4-velocity $u^\nu$. In the
Landau frame the fluid velocity is connected to the energy flow, which
implies ${q}^{\nu}=0$ or
$u_\mu {T}^{\mu\nu} = {\epsilon} u^\nu$~\cite{Landau1987}.  In the
Eckart frame the fluid velocity is connected to the particle flows,
\ie, ${j}^\mu_a=0$ or
${N}^\mu_a = {n}_a u^\mu$~\cite{1940PhRv...58..919E}. In our
derivations, we will maintain a generic fluid velocity without
anchoring it to a specific reference frame.

\subsection{Decomposing the thermodynamic force in different dissipative processes}
\label{sec:C_decomp1}

The averages of Eqs.~\eqref{eq:T_munu_decomp} and \eqref{eq:N_a_decomp} over
the relevant distribution can now be substituted into
Eq.~\eqref{eq:cons_laws}. As a result, one finds the equations of ideal
hydrodynamics, since the averages of the dissipative operators vanish in
local equilibrium~\cite{1979TMP....40..821Z}
\bea \label{eq:ideal_hydro2}
Dn_a+ n_a\theta =0,\qquad
D\epsilon+ h \theta =0,\qquad
h D u_\alpha = \nabla_\alpha p.
\eea
To account for dissipative phenomena, it is essential to examine the
statistical operator's departure from its local equilibrium
configuration.

To compute the statistical averages of the dissipative fluxes we
decompose the operator $\hat{C}$ given by Eq.~\eqref{eq:C_op} into
different dissipative processes, as it was done in
Ref.~\cite{Harutyunyan_hydro2022} and the previous
treatments~\cite{1979TMP....40..821Z,1984AnPhy.154..229H,2011AnPhy.326.3075H}. The
details of the computation are provided in \ref{app:C_decomp}, and
the final result reads
\bea \label{eq:C_decompose_2nd}
\hat{C}(x)=\hat{C}_1(x)+\hat{C}_2(x),
\eea
where $\hat{C}_1$ and $\hat{C}_2$ are the first- 
and the second-order contributions, respectively:
\bea\label{eq:op_C1} 
\hat{C}_1(x)&=&-\beta \theta \hat{p}^* +
\beta\hat{\pi}_{\rho\sigma}\sigma^{\rho\sigma}-
\sum\limits_a\hat{\mathscr J}^{\sigma}_a\nabla_\sigma\alpha_a,\\
\label{eq:op_C2}
\hat{C}_2(x)&=&-
\hat{\beta}^*\big(\Pi\theta +\partial_\mu q^{\mu}-
q^{\mu}\dot{u}_\mu -\pi^{\mu\nu}\sigma_{\mu\nu}\big)
+\sum\limits_a \hat{\alpha}_a^* (\partial_\mu j^{\mu}_a) \nonumber\\
&-& \hat{q}^{\sigma}\beta h^{-1}\big (-\nabla_\sigma \Pi
+\Pi \dot{u}_\sigma + \dot{q}_\sigma + q^{\mu}\partial_\mu u_{\sigma}
 +q_{\sigma}\theta +\Delta_{\sigma\nu}\partial_\mu \pi^{\mu\nu}\big),
\eea
where we denoted $\dot{u}_\sigma = Du_\sigma$,
$\dot{q}_\sigma = \Delta_{\sigma\nu} D q^{\nu}$.  It is natural to
refer to the expressions in parentheses in Eq.~\eqref{eq:op_C2} as
{\it generalized} or {\it extended thermodynamic forces}. These forces
are of the second order as they either involve space-time derivatives
of the dissipative fluxes or products of two first-order terms.
In Eq.~\eqref{eq:op_C1} the
operators
\bea\label{eq:diff_currents}
\mathscr{\hat{J}}_a^\sigma =\hat{j}^{\sigma}_a
-\frac{n_a}{h}\hat{q}^{\sigma}
\eea
are the diffusion fluxes, \ie, the charge currents with respect to the
energy flow, which are invariant under first-order changes in
$u^\mu$~\cite{1976AnPhy.100..310I,1979AnPhy.118..341I};
\bea\label{eq:p_star1}
\hat{p}^* = \hat{p} - \gamma\hat{\epsilon}
-\sum\limits_a\delta_a \hat{n}_a
\eea
is the operator of the bulk viscous pressure (see
Sec.~\ref{sec:2nd_bulk}), with the coefficients $\gamma$ and
$\delta_a$ given by
\bea\label{eq:gamma_delta_a1}
\gamma = \left.\frac{\partial p}{\partial\epsilon}
\right|_{n_a},\qquad
\delta_a = \left.\frac{\partial p}
{\partial n_a} \right|_{\epsilon, n_b\neq n_a}.
\eea
The operators $\hat{\beta}^*$ and $\hat{\alpha}^*_a$ in
Eq.~\eqref{eq:op_C2} are given by
\bea\label{eq:beta_star}
\hat{\beta}^* &=& \hat\epsilon\left.\frac{\partial\beta}
{\partial\epsilon}\right|_{n_a}
+\sum\limits_a\hat{n}_a\left.\frac{\partial \beta}
{\partial n_a} \right|_{\epsilon,n_b\neq n_a},\\
\label{eq:alpha_star}
\hat{\alpha}^*_a &=& \hat\epsilon\left.\frac{\partial \alpha_a}
{\partial\epsilon}\right|_{n_b}
+\sum\limits_c\hat{n}_c\left.\frac{\partial\alpha_a}
{\partial n_c}\right|_{\epsilon, n_b\neq n_c}.
\eea
Note that the thermodynamic force $\hat{C}$ enters the correlators in Eq.~\eqref{eq:stat_average} with the thermodynamic parameters
evaluated at the point $x_1$. As it was shown in
Ref.~\cite{Harutyunyan_hydro2022}, this induces non-local corrections
to the averages of the dissipative currents from the two-point
correlators which are of the second order.

In the following, we systematically derive these corrections and show
that additional terms arise due two-point quantum correlators between
tensors of different ranks, as these operators are evaluated at
distinct space-time points and a naive application of Curie's theorem
  does not hold. These were omitted in
Ref.~\cite{Harutyunyan_hydro2022}.  To compute the non-local
corrections we expand the thermodynamic forces
$\partial^{\mu}\beta^\nu$ and $\partial_\mu\alpha_a$ in the operator
$\hat{C}$ into series with respect to $x_1$ around $x$: here it is
more convenient to use the initial form~\eqref{eq:C_op} for the
operator $\hat{C}$. Keeping only the linear terms gives
$\partial^{\mu}\beta^\nu(x_1)=\partial^{\mu}\beta^\nu(x)+\partial_{\tau}\partial^{\mu}\beta^\nu(x)(x_1-x)^\tau$,
$\partial_\mu\alpha_a(x_1)=\partial_\mu\alpha_a(x)+
\partial_{\tau}\partial_\mu \alpha_a(x)(x_1-x)^\tau$, and
\bea\label{eq:C_expand}
\hat{C}(x_1) = \hat{C}(x_1)_x+\partial_\tau
 \hat{C}\cdot (x_1-x)^\tau,
\eea
where $\hat{C}(x_1)_x$ is obtained from $\hat{C}(x_1)$ via replacing
the arguments $x_1$ of all {\it hydrodynamic} quantities (but not the
microscopic quantum operators $\hat{T}_{\mu\nu}$ and $\hat{N}^\mu_a$)
with $x$. The computation of $\partial_\tau \hat{C}$ is provided
in~\ref{app:C_decomp}, and the final result is given by
Eq.~\eqref{eq:C_deriv1}.

Using Eqs.~\eqref{eq:C_decompose_2nd} and
\eqref{eq:C_expand} we can now write for the
operator $\hat{C}(x_1)$ 
\bea \label{eq:C_x1}
\hat{C}(x_1) =\hat{C}_1(x_1)_x+\hat{C}_2(x_1)_x
+\partial_\tau \hat{C}(x_1)_x\cdot (x_1-x)^\tau.
\eea
Now using Eqs.~\eqref{eq:stat_average} and 
\eqref{eq:C_x1} for the statistical average 
of any operator $\hat{X}(x)$ we can write
\bea\label{eq:stat_average_2nd}
\langle \hat{X}(x)\rangle 
= \langle \hat{X}(x)\rangle_l +
\langle \hat{X}(x)\rangle_1 +
\langle \hat{X}(x)\rangle_2.
\eea
The first-order correction is given by
\bea
\label{eq:stat_average_C1}
\langle \hat{X}(x)\rangle_1 =
\int\! d^4x_1 \Big(\hat{X}(x),\hat{C}_1(x_1)_x
\Big),
\eea
and the second-order correction $\langle \hat{X}(x)\rangle_2$ can be
written as a sum of three terms
\bea\label{eq:stat_average_2_new}
\langle \hat{X}(x)\rangle_2 =
\langle \hat{X}(x)\rangle_2^1
+\langle \hat{X}(x)\rangle_2^2
+\langle \hat{X}(x)\rangle_2^3,
\eea
where 
\bea\label{eq:stat_average_21}
\langle \hat{X}(x)\rangle_2^1 
&=& \int\! d^4x_1\Big(\hat{X}(x),
\partial_\tau \hat{C}(x_1)_x\Big)(x_1-x)^\tau,\\
\label{eq:stat_average_22}
\langle \hat{X}(x)\rangle_2^2 &=&
\int\! d^4x_1 \Big(\hat{X}(x),\hat{C}_2(x_1)_x\Big),\\
\label{eq:stat_average_23}
\langle \hat{X}(x)\rangle_2^3 &=&
\int\! d^4x_1d^4x_2
\Big(\hat{X}(x),\hat{C}_1(x_1)_x,\hat{C}_1(x_2)_x\Big).
\eea
Note that we omitted the second and third terms of Eq.~\eqref{eq:C_x1}
in Eq.~\eqref{eq:stat_average_23}, as they contribute only third-order
and higher corrections. The first term in
Eq.~\eqref{eq:stat_average_2_new} accounts for the non-local
corrections from the operator $\hat{C}(x_1)$. The second term
incorporates corrections from the generalized thermodynamic forces,
while the third term captures the quadratic corrections involving the
three thermodynamic forces $\theta$, $\sigma_{\rho\sigma}$, and
$\nabla_\sigma\alpha_a$.

It is useful to note that the expansion~\eqref{eq:stat_average_2nd} in
powers of the thermodynamic forces is equivalent to the expansion in
powers of the Knudsen number ${\rm Kn}=\lambda/L$, where  in strongly coupled systems, $\lambda$ has the meaning of the
mean correlation length, which is the analog of the particle mean free
path of weakly coupled systems. Indeed, the integrands in
Eqs.~\eqref{eq:stat_average_C1} and
\eqref{eq:stat_average_21}--\eqref{eq:stat_average_23} are mainly
concentrated within the range $|x_{1,2}-x|\lesssim \lambda$. As a
result, we can estimate $|\partial_\tau \hat{C}|\simeq |\hat{C}|/L$,
and, \eg, the ratio of the averages given by
Eqs.~\eqref{eq:stat_average_C1} and \eqref{eq:stat_average_21} will be
of the order of ${\rm Kn}$.
 
\section{Computing the dissipative quantities}
\label{sec:diss1}

Before performing the computation of the statistical averages of the
dissipative currents we discuss Curie's theorem, which states that in
isotropic medium the two-point correlation functions between operators
of different ranks
vanish~\cite{1963PhT....16e..70D,1963AmJPh..31..558D}. This statement
is true for any {\it microscopic} operators such as the
energy-momentum tensor or charge currents, which are well-defined
without any reference to the hydrodynamic regime for the
system. However, the theorem can also be applied to the correlators
between {\it macroscopic} or {\it hydrodynamic} operators given by
Eqs.~\eqref{eq:proj1_op} and \eqref{eq:proj2_op} in the case if the
difference between the fluid velocities, and, therefore, also the
projectors $\Delta_{\alpha\beta}$ at the points $x$ and $x_1$ can be
neglected, \ie, if we need first-order accuracy only. However, if
second-order terms are of interest as well, this theorem requires an
  extension. Specifically, the presence of the acceleration vector in
  the system allows for building invariants from tensors of different
  ranks.

To make the argument clearer we give an explicit example. The theorem
works for the correlator
$\Big(\hat{\pi}_{\mu\nu}(x),\mathscr{\hat{J}}_b^{\sigma}(x_1)_x\Big)$
which vanishes as the operators $\hat{\pi}_{\mu\nu}(x)$ and
$\mathscr{\hat{J}}_b^{\sigma}(x_1)_x$ are obtained by projecting
$\hat{T}^{\mu\nu}$ and $\hat{N}^{\mu}_a$ with {\it the same projector}
$\Delta_{\alpha\beta}$ which is evaluated at the point $x$. On the
other hand, the correlator
$\Big(\hat{\pi}_{\mu\nu}(x),\mathscr{\hat{J}}_b^{\sigma}(x_1)\Big)$
does not vanish, as the operators $\hat{\pi}_{\mu\nu}(x)$ and
$\mathscr{\hat{J}}_b^{\sigma}(x_1)$ contain two different projectors,
\ie, $\Delta_{\alpha\beta}(x)$ and $\Delta_{\alpha\beta}(x_1)$. The
physical reason why the difference between these two correlators
\bea 
\Big(\hat{\pi}_{\mu\nu}(x),\mathscr{\hat{J}}_b^{\sigma}(x_1)\Big)
-
\Big(\hat{\pi}_{\mu\nu}(x),\mathscr{\hat{J}}_b^{\sigma}(x_1)_x\Big)\simeq
\Big(\hat{\pi}_{\mu\nu}(x),\partial_\tau \mathscr{\hat{J}}_b^{\sigma}(x_1)_x\Big)
(x_1-x)^\tau
\eea
can be nonzero is the fact that the derivative
$\partial_\tau \mathscr{\hat{J}}_b^{\sigma}$ contains a term
$\propto \hat{\pi}_{\tau}^{\sigma}(x_1)_x$ which then couples to
$\hat{\pi}_{\mu\nu}(x)$. In other words, the non-locality of the
thermodynamic force which is taken into account by the formal
derivative $\partial_\tau\hat{C}$ in Eq.~\eqref{eq:stat_average_21},
produces mixing between dissipative phenomena of different ranks, which
leads to additional non-local terms in the transport equations. These
mixed terms were omitted in our previous
paper~\cite{Harutyunyan_hydro2022}, where, \eg, in the case of shear
stresses only the non-local term
$\Big(\hat{\pi}_{\mu\nu}(x),\partial_\tau
\hat{\pi}_{\rho\sigma}(x_1)_x\Big)$ was kept.

As these new terms arise because of the velocity gradients, they are
always proportional to the fluid acceleration $\dot{u}_\mu$, as shown
below.

\subsection{First-order averages}

Substituting Eq.~\eqref{eq:op_C1} into Eqs.~\eqref{eq:stat_average_C1}
and taking into account Curie's theorem (note that in the operator
$\partial_\tau \hat{C}(x_1)_x$, all velocities are evaluated at $x$,
allowing Curie's theorem to be applied without issue) we can compute
the first-order corrections to the shear stress tensor, the bulk
viscous pressure and the diffusion currents~\footnote{See
  Sec.~\ref{sec:2nd_bulk} for details regarding the bulk viscous
  pressure.}
\bea \label{eq:stress_av2}
\langle \hat{\pi}_{\mu\nu}(x)\rangle_1
&=&\beta(x) \sigma^{\rho\sigma}(x)\!\int d^4x_1
\Big(\hat{\pi}_{\mu\nu}(x),\hat{\pi}_{\rho\sigma}
(x_1)_x\Big),\\
\label{eq:pressure_av} 
\langle \hat{p}^*(x)\rangle_1 &=& -\beta(x) \theta(x) \!
\int\! d^4x_1 \Big(\hat{p}^* (x),\hat{p}^*(x_1)_x\Big),\\
\label{eq:charge_currents}
\langle\mathscr{\hat{J}}_a^{\mu}(x)\rangle_1 
&=& - \sum\limits_b\nabla_\sigma\alpha_b(x) \!
\int d^4x_1
\left(\mathscr{\hat{J}}_a^{\mu}(x),
\mathscr{\hat{J}}_b^{\sigma}(x_1)_x\right).
\eea
Equations \eqref{eq:stress_av2}--\eqref{eq:charge_currents} establish
the required linear relations between the dissipative fluxes and the
thermodynamic forces. The isotropy of the medium together with the
conditions \eqref{eq:orthogonality} further
implies~\cite{1984AnPhy.154..229H}
\bea \label{eq:corr1_current}
\Big(\mathscr{\hat{J}}_a^{\mu}( x),\mathscr{\hat{J}}_b^{\sigma}(x_1)_x\Big)& = & 
\frac{1}{3}\Delta^{\mu\sigma}(x)\Big(\mathscr{\hat{J}}_a^{\lambda}(x),
\mathscr{\hat{J}}_{b\lambda}(x_1)_x\Big),\\
\label{eq:corr1_stress}
\Big(\hat{\pi}_{\mu\nu}(x),
\hat{\pi}_{\rho\sigma}( x_1)_x\Big)
 & =& \frac{1}{5} \Delta_{\mu\nu\rho\sigma}(x)
\Big(\hat{\pi}^{\lambda\eta}(x),
\hat{\pi}_{\lambda\eta}( x_1)_x\Big).
\eea
Defining the shear and the bulk viscosities and the matrix of
diffusion coefficients as
\bea \label{eq:shear_def}
\eta(x) &=& \frac{\beta(x)}{10}\int d^4x_1
\Big(\hat{\pi}_{\mu\nu}(x),\hat{\pi}^{\mu\nu}(x_1)_x\Big),\\
 \label{eq:bulk_def}
\zeta(x) &=& \beta(x)\int d^4x_1
\Big(\hat{p}^*(x), \hat{p}^*(x_1)_x\Big),\\
\label{eq:chi_ab_def}
\chi_{ab}(x) &=& -\frac{1}{3}\int d^4x_1
\left(\mathscr{\hat{J}}_a^{\lambda}
(x),\mathscr{\hat{J}}_{b\lambda}(x_1)_x\right),
\eea
we obtain from Eqs.~\eqref{eq:stress_av2}--\eqref{eq:chi_ab_def} 
\bea \label{eq:shear_bulk_diff}
\langle \hat{\pi}_{\mu\nu}\rangle_1 =
 2\eta \sigma_{\mu\nu},\qquad
\langle \hat{p}^*\rangle_1
=-\zeta\theta,\qquad
\langle\hat{\mathscr{J}}_a^\mu\rangle_1
= \sum\limits_b\chi_{ab}\nabla^\mu \alpha_b.
\eea
In the case of one sort of conserved charge 
we define the thermal conductivity as
\bea\label{eq:kappa_def}
\kappa =-\frac{\beta^2}{3}\int d^4x_1
\left(\hat{h}^{\lambda}(x),
\hat{h}_{\lambda}(x_1)_x\right),
\eea
where we defined the heat-flux operator as
\bea\label{eq:op_heat_flux}
\hat{h}^\lambda = \hat{q}^{\lambda}
-\frac{h}{n}\hat{j}^{\lambda}=-\frac{h}{n}\mathscr{\hat{J}}^{\lambda}.
\eea

The two-point correlators in
Eqs.~\eqref{eq:shear_def}--\eqref{eq:chi_ab_def} can be expressed via
two-point retarded Green's functions (see ~\ref{app:Green_func})
\bea \label{eq:shear_bulk_mod}
\eta = -\frac{1}{10}\frac{d}{d\omega} {\rm Im}
G^R_{\hat{\pi}_{\mu\nu}\hat{\pi}^{\mu\nu}}(\omega)
\bigg\vert_{\omega=0},\qquad
\zeta = -\frac{d}{d\omega} {\rm Im}G^R_{\hat{p}^{*}
\hat{p}^*}(\omega)\bigg\vert_{\omega=0},\\
\label{eq:kappa_mod}
\chi_{ab} = \frac{T}{3}\frac{d}{d\omega} {\rm Im}
G^R_{\hat{\mathscr{J}}_a^{\lambda}\hat{\mathscr{J}}_{b\lambda}}(\omega)
\bigg\vert_{\omega=0},\qquad
\kappa = \frac{1}{3T}\frac{d}{d\omega} {\rm Im}
G^R_{\hat{h}^{\lambda}\hat{h}_{\lambda}}(\omega)
\bigg\vert_{\omega=0},
\eea
where 
\bea\label{eq:green_func}
G^R_{\hat{X}\hat{Y}}(\omega) 
= -i\int_{0}^{\infty} dt
e^{i\omega t}\int d^3x\langle\big[\hat{X}(\bm x, t),
\hat{Y}(\bm 0,0)\big]\rangle_l
\eea
is the Fourier transform of the two-point retarded correlator taken in
the zero-wavenumber limit and the square brackets denote the
commutator. Relations \eqref{eq:shear_bulk_mod} and
\eqref{eq:kappa_mod} are known as Kubo
formulas~\cite{1957JPSJ...12..570K,
  1957JPSJ...12.1203K,1984AnPhy.154..229H,2011AnPhy.326.3075H,1987NuPhB.280..716H}.

\subsection{Second-order corrections to the shear stress tensor}
\label{sec:2nd_shear}

By substituting Eq.~\eqref{eq:C_deriv1} into
Eq.~\eqref{eq:stat_average_21} and using Eq.~\eqref{eq:corr1_stress},
we obtain the result
\bea\label{eq:shear_av_non_local1}
\langle \hat{\pi}_{\mu\nu}(x)\rangle_2^1  
&=& 2\beta^{-1}(x)\Delta_{\mu\nu\rho\sigma}(x)\!
\left[\partial_{\tau} (\beta \sigma^{\rho\sigma})
+(\partial_{\tau}u^\rho) \sum_a \frac{n_a}{h}\nabla^\sigma\alpha_a\right]_x a^\tau,
\eea
where
\bea
a^\tau = \frac{\beta(x)}{10} 
\int\! d^4x_1\Big(\hat{\pi}^{\mu\nu}(x),
\hat{\pi}_{\mu\nu}(x_1)_x\Big)(x_1-x)^\tau.
\eea
Here we substituted the two-point correlation function given by
Eqs.~\eqref{eq:corr1_stress} and factored out the thermodynamic force
in the square brackets from the integral. The last term in square
brackets is the one which arises from the mixing between the shear
stresses and the diffusion currents, which were omitted in the
previous treatment.

According to Eqs.~\eqref{eq:vector_I0_rest}--\eqref{eq:vec_gen}
in~\ref{app:Green_func} the vector $a^\tau$ can be 
written in the following form 
\bea \label{eq:a_tau}
a^\tau =-\eta\tau_\pi u^\tau,
\eea
where we defined
\bea\label{eq:eta_relax}
\eta\tau_\pi = -i\frac{d}{d\omega}\eta(\omega)\bigg\vert_{\omega=0}=
\frac{1}{20}\frac{d^2}{d\omega^2} {\rm Re}G^R_{\hat{\pi}_{ij}
\hat{\pi}^{ij}(\omega)}\bigg\vert_{\omega=0}.
\eea
Here the retarded Green's function is given by
Eq.~\eqref{eq:green_func}, and the frequency-dependent shear viscosity
$\eta(\omega)$ is defined by analogy with Eq.~\eqref{eq:I_XY}. As seen
from Eq.~\eqref{eq:eta_relax}, the new coefficient $\tau_\pi$ has a
dimension of time and can be regarded as a relaxation time for the
shear stress tensor. Combining Eqs.~\eqref{eq:shear_av_non_local1} and
\eqref{eq:a_tau} we obtain
\bea\label{eq:shear_av_21}
\langle \hat{\pi}_{\mu\nu}\rangle_2^1 
&=& - 2\eta\tau_\pi \beta^{-1}\Delta_{\mu\nu\rho\sigma}
D (\beta \sigma^{\rho\sigma}) 
- 2\eta\tau_\pi \beta^{-1}
\Delta_{\mu\nu\rho\sigma} D u^\rho \sum_a 
\frac{n_a}{h}\nabla^\sigma\alpha_a\nonumber\\
&=& -2\eta\tau_\pi (\Delta_{\mu\nu\rho\sigma} 
D\sigma^{\rho\sigma}+\gamma\theta \sigma_{\mu\nu}) 
- 2\eta\tau_\pi Th^{-1}\sum_a n_a
\dot{u}_{<\mu} \nabla_{\nu >}\alpha_a,
\eea
where we used Eq.~\eqref{eq:D_beta1} in the second step (keeping only
the leading-order term) and omitted the argument of
$\hat{\pi}_{\mu\nu}$ for brevity. The last terms in each of the
expressions on the right-hand side are new.

The averages~\eqref{eq:stat_average_22} and \eqref{eq:stat_average_23}
for $\hat{\pi}_{\mu\nu}$ were computed in
Ref.~\cite{Harutyunyan_hydro2022}
\bea\label{eq:shear_av_22}
\langle \hat{\pi}_{\mu\nu}\rangle_2^2 &=& 0,\\
\label{eq:shear_av_23}
\langle \hat{\pi}_{\mu\nu}\rangle_2^3 
&=& \lambda_\pi \sigma_{\alpha<\mu}\sigma_{\nu>}^{\alpha}+
2\lambda_{\pi\Pi}\theta \sigma_{\mu\nu}
+\sum\limits_{ab}\lambda_{\pi{\!\mathscr J}}^{ab}
\nabla_{<\mu}\alpha_a\nabla_{\nu>}\alpha_b,
\eea
with the coefficients
\bea\label{eq:lambda_pi}
\lambda_\pi &=& \frac{12}{35}\beta^2\! \int d^4x_1d^4x_2
\Big(\hat{\pi}_{\gamma}^{\delta}(x),\hat{\pi}_{\delta}^{\lambda}(x_1)_x,
\hat{\pi}_{\lambda}^{\gamma}(x_2)_x\Big),\\
\label{eq:lambda_2}
\lambda_{\pi\Pi} &=& -\frac{\beta^2 }{5}\! \int d^4x_1d^4x_2
\Big(\hat{\pi}_{\gamma\delta}(x),\hat{\pi}^{\gamma\delta}(x_1)_x,
\hat{p}^*(x_2)_x\Big),\\
\label{eq:lambda1_ab}
\lambda_{\pi{\!\mathscr J}}^{ab} &=& 
\frac{1}{5}\!\int d^4x_1d^4x_2
\Big(\hat{\pi}_{\gamma\delta}(x),
\hat{\mathscr J}^{\gamma}_a(x_1)_x,
\hat{\mathscr J}^{\delta}_b(x_2)_x\Big).
\eea

\subsubsection{Final equation for the shear stress tensor}
\label{sec:final_shear}

Combining all corrections from Eqs.~\eqref{eq:shear_bulk_diff},
\eqref{eq:shear_av_21}, \eqref{eq:shear_av_22} and
\eqref{eq:shear_av_23} and using Eqs.~\eqref{eq:stat_average_2nd} and
\eqref{eq:stat_average_2_new} we obtain the complete second-order
expression for the shear stress tensor
\bea\label{eq:shear_total_final}
{\pi}_{\mu\nu}
&=&  2\eta \sigma_{\mu\nu}
 -2\eta\tau_\pi (\Delta_{\mu\nu\rho\sigma}D \sigma^{\rho\sigma}+ 
 \gamma \theta \sigma_{\mu\nu})
 - 2\eta\tau_\pi Th^{-1}\sum_a n_a
\dot{u}_{<\mu} \nabla_{\nu >}\alpha_a \nonumber\\
 &&  +\lambda_\pi \sigma_{\alpha<\mu}\sigma_{\nu>}^{\alpha} +
  2\lambda_{\pi\Pi}\theta \sigma_{\mu\nu}
 +  \sum\limits_{ab}\lambda_{\pi{\!\mathscr J}}^{ab} 
 \nabla_{<\mu}\alpha_a\nabla_{\nu>}\alpha_b.
\eea
Here the second-order terms in the first line represent the non-local
corrections, whereas the second line collects the nonlinear
corrections from the three-point correlations.

We then modify Eq.~\eqref{eq:shear_total_final} to derive a
relaxation-type equation for ${\pi}_{\mu\nu}$ by replacing
$2\sigma^{\rho\sigma}$ with $\eta^{-1}{\pi}^{\rho\sigma}$ in the
second term of the right-hand side of Eq.~\eqref{eq:shear_total_final}
as was also done in Ref.~\cite{Harutyunyan_hydro2022} and previously
suggested in
Refs.~\cite{2008JHEP...04..100B,2013PhRvC..87e1901J,2015JHEP...02..051F}.
Such substitution is justified because this term is of the second
order in the space-time gradients. We then have
\bea\label{eq:shear_relax}
-2\eta\tau_\pi \Delta_{\mu\nu\rho\sigma}D \sigma^{\rho\sigma}
\simeq  -\tau_\pi \dot{\pi}_{\mu\nu}+ \tau_\pi\beta\eta^{-1} 
\bigg(\gamma \frac{\partial\eta}{\partial \beta}
-\sum\limits_a\delta_a \frac{\partial\eta}
{\partial \alpha_a}\bigg)\theta {\pi}_{\mu\nu}, 
\eea
where we defined $\dot{\pi}_{\mu\nu}=\Delta_{\mu\nu\rho\sigma}
D{\pi}^{\rho\sigma}$ and used Eqs.~\eqref{eq:D_beta1} and \eqref{eq:D_alpha1}
at the leading order. Combining Eqs.~\eqref{eq:shear_total_final} 
and \eqref{eq:shear_relax} and introducing the coefficients
\bea\label{eq:lambda}
\lambda &=& 2(\lambda_{\pi\Pi}-\gamma\eta\tau_\pi),\\
\label{eq:tilde_lambda_pi}
\tilde{\lambda}_\pi &=& \tau_\pi \beta \eta^{-1}
\bigg(\gamma \frac{\partial\eta}{\partial \beta}
-\sum\limits_a\delta_a \frac{\partial\eta}{\partial \alpha_a}\bigg),
\eea
we obtain finally
\bea\label{eq:shear_final_relax}
{\pi}_{\mu\nu}
= 2\eta \sigma_{\mu\nu}-\tau_\pi \dot{\pi}_{\mu\nu}
+\tilde{\lambda}_\pi\theta\pi_{\mu\nu}
 - 2\eta\tau_\pi Th^{-1}\sum_a n_a
\dot{u}_{<\mu} \nabla_{\nu >}\alpha_a \nonumber\\
+ \lambda_\pi  \sigma_{\alpha<\mu}\sigma_{\nu>}^{\alpha}
 + \lambda \theta {\sigma}_{\mu\nu} +
 \sum\limits_{ab}\lambda_{\pi{\!\mathscr J}}^{ab}
 \nabla_{<\mu}\alpha_a\nabla_{\nu>}\alpha_b.
\eea
Here the fourth term on the right-hand side is new.

\subsection{Second-order corrections to the bulk viscous pressure}
\label{sec:2nd_bulk}

As is known, the bulk viscous pressure measures the deviation 
of the actual thermodynamic pressure $\langle\hat{p}\rangle$
from its equilibrium value $p(\epsilon,n_a)$ given by the EoS 
as a result of fluid expansion or compression
\bea\label{eq:bulk}
\Pi = \langle\hat{p}\rangle-{p}(\epsilon,n_a)=
\langle\hat{p}\rangle_l +\langle\hat{p}\rangle_1
+\langle\hat{p}\rangle_2 -{p}(\epsilon,n_a).
\eea
Taking into account the possible deviations of the energy 
and charge densities from their equilibrium values 
$\epsilon=\langle\hat{\epsilon}\rangle_l +\Delta\epsilon$, 
$n_a=\langle\hat{n}_a\rangle_l +\Delta n_a$ we obtain
\bea\label{eq:pressure_expand1}
\langle\hat{p}\rangle_l \equiv
{p}\big(\langle\hat{\epsilon}\rangle_l,
\langle\hat{n}_a\rangle_l\big) = {p}\big({\epsilon}-
\Delta{\epsilon},{n}_a-\Delta{n}_a \big) =
p({\epsilon},{n}_a)-\gamma\Delta\epsilon -
\sum\limits_a\delta_a\Delta n_a \nonumber\\
+\psi_{\epsilon\epsilon}\Delta\epsilon^2 + 
2\sum\limits_a\psi_{\epsilon a} \Delta\epsilon\Delta n_a+
\sum\limits_{ab}\psi_{ab} \Delta n_a\Delta n_b,
\eea
where the coefficients $\gamma$, $\delta_a$ are defined in
Eq.~\eqref{eq:gamma_delta_a1}, and 
\bea\label{eq:xi_ab}
\psi_{\epsilon\epsilon} =\frac{1}{2}
\frac{\partial^2 p}{\partial\epsilon^2},\qquad
\psi_{\epsilon a} =\frac{1}{2} \frac{\partial^2 p}
{\partial\epsilon\partial n_a},\qquad
\psi_{ab} =\frac{1}{2} \frac{\partial^2 p}
{\partial n_a\partial n_b}.
\eea
Note that the corrections
$\Delta\epsilon =\langle\hat{\epsilon}\rangle_1
+\langle\hat{\epsilon}\rangle_2$ and
$\Delta n_a = \langle\hat{n}_a\rangle_1+\langle\hat{n}_a\rangle_2$
vanish if the matching conditions~\eqref{eq:matching} are imposed.  We
maintain these parameters to preserve the generality of the
expressions, ensuring they remain independent of specific matching
condition selections. By substituting Eq.~\eqref{eq:pressure_expand1}
in Eq.~\eqref{eq:bulk} as well as $\Delta\epsilon$ and $\Delta n_a$,
and focusing exclusively on second-order terms for bulk viscous
pressure, we derive the following result
\bea\label{eq:bulk_2}
\Pi =
\langle\hat{p}^*\rangle_1+\langle\hat{p}^*\rangle_2
+\psi_{\epsilon\epsilon}\langle\hat{\epsilon}\rangle_1^2+
2\sum\limits_a\psi_{\epsilon a} \langle\hat{\epsilon}\rangle_1
\langle\hat{n}_a\rangle_1 +\sum\limits_{ab}\psi_{ab} 
\langle\hat{n}_a\rangle_1 \langle\hat{n}_b\rangle_1,
\eea
where we used the definition \eqref{eq:p_star1} of $\hat{p}^*$.
Upon introducing the coefficients [see Eq.~\eqref{eq:I_XY5}]
\bea\label{eq:zeta_ep}
\zeta_\epsilon &=& \beta\! \int\! d^4x_1
\Big(\hat{\epsilon}(x),\hat{p}^*(x_1)_x\Big)
=-\frac{d}{d\omega} {\rm Im}G^R_{\hat{\epsilon}
\hat{p}^*}(\omega)\bigg\vert_{\omega=0},\\
\label{eq:zeta_n_a}
\zeta_{a} &=& \beta\! \int\! d^4x_1
\Big(\hat{n}_a(x),\hat{p}^*(x_1)_x\Big)=
-\frac{d}{d\omega} {\rm Im}G^R_{\hat{n}_a
\hat{p}^*}(\omega)\bigg\vert_{\omega=0},
\eea
according to Eqs.~\eqref{eq:op_C1} and \eqref{eq:stat_average_C1} 
the averages $\langle\hat{\epsilon}\rangle_1$ and
$\langle\hat{n}_a\rangle_1$ can be written as
\bea\label{eq:ep_n_1st_order}
\langle \hat{\epsilon}\rangle_1
= -\zeta_\epsilon\theta,\qquad
\langle \hat{n}_a\rangle_1
= -\zeta_{a}\theta.
\eea
Then we have from Eqs.~\eqref{eq:shear_bulk_diff}, \eqref{eq:bulk_2}, and
\eqref{eq:ep_n_1st_order}
\bea\label{eq:bulk_sum}
\Pi =-\zeta\theta 
+\Big(\psi_{\epsilon\epsilon}\zeta_\epsilon^2+
2\zeta_\epsilon \sum\limits_a\psi_{\epsilon a}  \zeta_{a} +
 \sum\limits_{ab}\psi_{ab} \zeta_{a} \zeta_{b} \Big)\theta^2
+\langle\hat{p}^*\rangle_2.
\eea

Next we compute $\langle\hat{p}^*\rangle_2$. Using Eqs.~\eqref{eq:C_deriv1} and \eqref{eq:stat_average_21}
and Curie's theorem we obtain
\bea\label{eq:bulk_av_non_local1}
\langle \hat{p}^*(x)\rangle_2^1 
&=& -\partial_{\tau}(\beta\theta)\! \int\! d^4x_1
\Big(\hat{p}^*(x),\hat{p}^*(x_1)_x\Big)(x_1-x)^\tau \nonumber\\
&& +\beta\theta\! \int\! d^4x_1\Big(\hat{p}^*(x),
\Big[\hat{\epsilon} (\partial_{\tau}\gamma) + \sum\limits_a \hat{n}_a 
(\partial_{\tau}\delta_a) \Big] \Big)(x_1-x)^\tau\nonumber\\
&& -\sum_a (\partial_{\tau}u_{\rho}) 
\left(\nabla^\rho\alpha_a\right)
\int\! d^4x_1\Big(\hat{p}^*(x),\left[n_a h^{-1}(\hat{\epsilon}+\hat{p}) - \hat{n}_a\right]\Big)(x_1-x)^\tau,
\eea
where the integrals can be expressed as
\bea
&&\int\! d^4x_1
\Big(\hat{p}^*(x),\hat{p}^*(x_1)_x\Big)(x_1-x)^\tau
= -u^\tau \beta^{-1}\zeta\tau_\Pi, \\
&&\int\! d^4x_1\Big(\hat{p}^*(x),
\Big[\hat{\epsilon} (\partial_{\tau}\gamma) + \sum\limits_a \hat{n}_a 
(\partial_{\tau}\delta_a) \Big] \Big)(x_1-x)^\tau \nonumber\\
&& \hspace{2.5cm}=
-\beta^{-1}\left(\zeta_\epsilon\tau_\epsilon D\gamma+
  \sum\limits_a \zeta_{a}\tau_a D\delta_{a}\right),\\
&&\int\! d^4x_1
\Big(\hat{p}^*(x),\left[n_a h^{-1}(\hat{\epsilon}+\hat{p}) -
\hat{n}_a\right]\Big)(x_1-x)^\tau= -\beta^{-1}u^\tau \nonumber\\
&&\hspace{0.5cm}\left[n_a h^{-1} \zeta \tau_\Pi+ (1+\gamma)n_a h^{-1} 
\zeta_\epsilon\tau_\epsilon + \sum\limits_b \zeta_{b}
\tau_b (n_a \delta_{b}h^{-1}-\delta_{ab})\right].
\eea
Using Eqs.~\eqref{eq:vector_I0_rest}--\eqref{eq:vec_gen} 
in~\ref{app:Green_func} and $D(\beta\theta) \beta^{-1} =D\theta +\gamma\theta^2$
we obtain 
\bea\label{eq:bulk_av_non_local2}
\langle \hat{p}^*\rangle_2^1 &=& (D\theta +\gamma\theta^2)\zeta\tau_\Pi 
- \theta \left(\zeta_\epsilon\tau_\epsilon D\gamma+
 \sum\limits_a \zeta_{a}\tau_a D\delta_{a}\right)
+ \beta^{-1}\dot{u}_{\rho}\sum_a
 \left(\nabla^\rho\alpha_a\right) 
\nonumber\\
&&
\times\left[n_a h^{-1} \zeta \tau_\Pi+ (1+\gamma)n_a h^{-1} \zeta_\epsilon\tau_\epsilon + \sum\limits_b \zeta_{b}
\tau_b (n_a \delta_{b}h^{-1}-\delta_{ab})\right],
\eea
where $\delta_{ab}$ is the Kronecker delta, and the new 
coefficients $\tau_\Pi$, $\tau_\epsilon$ and $\tau_a$
are given by 
\bea\label{eq:zeta_relax}
\zeta \tau_\Pi &=& -i\frac{d}{d\omega}
\zeta(\omega) \bigg\vert_{\omega=0}=
\frac{d^2}{d\omega^2} {\rm Re}G^R_{\hat{p}^*
\hat{p}^*}(\omega)
\bigg\vert_{\omega=0},\\
\label{eq:zeta_tilde_ep}
\zeta_\epsilon \tau_\epsilon &=&
-i\frac{d}{d\omega}
\zeta_\epsilon(\omega) \bigg\vert_{\omega=0}= 
\frac{d^2}{d\omega^2} {\rm Re}G^R_{\hat{p}^*
\hat{\epsilon}}(\omega)
\bigg\vert_{\omega=0},\\
\label{eq:zeta_tilde_a}
\zeta_{a} \tau_a &=& -i\frac{d}{d\omega}
\zeta_{a}(\omega) \bigg\vert_{\omega=0}=
\frac{d^2}{d\omega^2} {\rm Re}G^R_{\hat{p}^*
\hat{n}_a}(\omega)
\bigg\vert_{\omega=0},
\eea
where $\zeta$, $\zeta_\epsilon$ and $\zeta_a$ in the limit $\omega\to 0$ are defined in Eqs.~\eqref{eq:bulk_def}, 
\eqref{eq:zeta_ep} and \eqref{eq:zeta_n_a}, respectively. 
In the case of $\omega\neq 0$ the formula \eqref{eq:I_XY} 
should be used with the relevant choices of the operators $\hat{X}$ and $\hat{Y}$. 
The last line in Eq.~\eqref{eq:bulk_av_non_local2} collects the new
terms which account for the non-local mixing between the bulk viscous
pressure and the diffusion currents. 

Next, using the definitions in Eq.~\eqref{eq:xi_ab}
we can write 
\bea
D\gamma &=& 2\Big(\psi_{\epsilon\epsilon} 
D\epsilon +\sum\limits_a \psi_{\epsilon a} D n_a\Big)
=-2\Big(\psi_{\epsilon\epsilon}h\theta +\theta
\sum\limits_a \psi_{\epsilon a} n_a\Big),\\
D\delta_a &=& 2\Big (\psi_{\epsilon a} 
D\epsilon +\sum\limits_{b} \psi_{ab} D n_{b}\Big)
=-2\Big (\psi_{\epsilon a} h\theta +\theta
\sum\limits_{b} \psi_{ab} n_b\Big),
\eea
where the derivatives $D\epsilon$ and $Dn_a$ were
eliminated by employing Eq.~\eqref{eq:ideal_hydro2}. Denoting
\bea\label{eq:zeta_star0}
{\zeta}^* &=& \gamma\zeta\tau_\Pi +2\zeta_\epsilon 
\tau_\epsilon\Big(\psi_{\epsilon\epsilon} h +
\sum\limits_a n_a\psi_{\epsilon a}\Big) +
2\sum\limits_a\zeta_{a}\tau_a\Big (\psi_{\epsilon a} h 
+\sum\limits_{b} \psi_{ab}  n_{b}\Big),\\
\label{eq:zeta_bar_a}
\bar{\zeta}_a &=& Tn_a h^{-1} \Big[\zeta \tau_\Pi+ 
(1+\gamma)\zeta_\epsilon\tau_\epsilon\Big] + 
T\sum\limits_b \zeta_{b} \tau_b (n_a \delta_{b}h^{-1}-\delta_{ab}),
\eea
we obtain for Eqs.~\eqref{eq:bulk_av_non_local2}
\bea\label{eq:bulk_av_21}
\langle \hat{p}^*\rangle_2^1 
= \zeta \tau_\Pi D\theta +\zeta^*\theta^2
+\sum_a \bar{\zeta}_a \dot{u}_{\rho} \nabla^\rho\alpha_a,
\eea
where the last term is new.

For the corrections $\langle \hat{p}^*\rangle_2^2$ and 
$\langle \hat{p}^*\rangle_2^3$ we have~\cite{Harutyunyan_hydro2022}
($\dot{u}_\mu=Du_\mu$)
\bea\label{eq:bulk_av_22}
\langle \hat{p}^*\rangle_2^2 &=&
\sum\limits_a \zeta_{\alpha_a}
\partial_\mu {\mathscr J}^{\mu}_a 
-\zeta_\beta (\Pi\theta -\pi^{\mu\nu}\sigma_{\mu\nu})
-\tilde{\zeta}_\beta \partial_\mu q^{\mu} \nonumber\\
&& \hspace{1.75cm} +q^\mu \Big[\zeta_\beta \dot{u}_\mu 
+\sum\limits_a \zeta_{\alpha_a} 
\nabla_\mu (n_ah^{-1})\Big],\qquad\\
\label{eq:bulk_av_23}
\langle \hat{p}^*\rangle_2^3 &=&
\lambda_\Pi \theta^2
-\lambda_{\Pi\pi} \sigma_{\alpha\beta}\sigma^{\alpha\beta} +
T\sum\limits_{ab}\zeta^{ab}_\Pi
\nabla^\sigma\alpha_a\nabla_\sigma\alpha_b.
\eea
In Eq.~\eqref{eq:bulk_av_22} we defined new transport coefficients by
\bea\label{eq:zeta_beta}
\zeta_\beta &=& \int\! d^4x_1
\Big(\hat{p}^*(x),\hat{\beta}^*(x_1)_x\Big)
=T \frac{\partial\beta} {\partial\epsilon} \,
\zeta_\epsilon +\sum\limits_c  T
\frac{\partial \beta}{\partial n_c}\, \zeta_c,\\
\label{eq:zeta_alpha_a}
\zeta_{\alpha_a} &=& \int\! d^4x_1 
\Big(\hat{p}^*(x),\hat{\alpha}_a^*(x_1)_x\Big)
=T \frac{\partial \alpha_a}{\partial\epsilon}\,
\zeta_\epsilon  +\sum\limits_c  T 
\frac{\partial\alpha_a} {\partial n_c}\,\zeta_c\,.
\eea
where we used Eqs.~\eqref{eq:beta_star}, \eqref{eq:alpha_star}, 
\eqref{eq:zeta_ep} and \eqref{eq:zeta_n_a} respectively, and
\bea\label{eq:tilde_zeta_beta}
\tilde{\zeta}_\beta ={\zeta}_\beta-h^{-1}
\sum\limits_a n_a\zeta_{\alpha_a}.
\eea
The coefficients in Eq.~\eqref{eq:bulk_av_23} are given by
\bea\label{eq:lambda_Pi}
\lambda_\Pi &=& \beta^2\!\int\! d^4x_1d^4x_2
\Big(\hat{p}^*(x),\hat{p}^*(x_1)_x,\hat{p}^*(x_2)_x\Big),\\
\label{eq:lambda_1}
\lambda_{\Pi\pi} &=& -\frac{\beta^2 }{5}\!\int\! d^4x_1d^4x_2
\Big(\hat{p}^*(x),\hat{\pi}_{\gamma\delta}(x_1)_x,
\hat{\pi}^{\gamma\delta}(x_2)_x\Big),\\
\label{eq:zeta1_ab}
\zeta^{ab}_\Pi &=& \frac{\beta}{3}\!\int\! 
d^4x_1d^4x_2 \Big(\hat{p}^*(x),
\hat{\mathscr J}_{a\gamma}(x_1)_x,
\hat{\mathscr J}^{\gamma}_b(x_2)_x\Big).
\eea

\subsubsection{Final equation for the bulk viscous pressure}
\label{sec:final_bulk}

Combining all pieces from Eqs.~\eqref{eq:bulk_av_21}, 
\eqref{eq:bulk_av_22} and \eqref{eq:bulk_av_23} we obtain according to Eq.~\eqref{eq:stat_average_2_new}
\bea\label{eq:bulk_av_2_final}
\langle \hat{p}^*\rangle_2 &=&
\zeta\tau_\Pi D\theta -\zeta_\beta
(\Pi\theta -\pi^{\mu\nu}\sigma_{\mu\nu})
-\tilde{\zeta}_\beta \partial_\mu q^{\mu}
+(\lambda_\Pi +\zeta^*)\theta^2 \nonumber\\
&&
-\lambda_{\Pi\pi} \sigma_{\alpha\beta}\sigma^{\alpha\beta} 
+\sum_a \bar{\zeta}_a
\dot{u}_{\rho} \nabla^\rho\alpha_a 
+\sum\limits_a \zeta_{\alpha_a} \partial_\mu {\mathscr J}^{\mu}_a \nonumber\\
&&
 +T\sum\limits_{ab}\zeta^{ab}_\Pi
\nabla^\sigma\alpha_a\nabla_\sigma\alpha_b +q^\mu 
\Big[\zeta_\beta \dot{u}_\mu +\sum\limits_a \zeta_{\alpha_a} 
\nabla_\mu (n_ah^{-1})\Big],\quad
\eea
where the second term on the second line is new.

To derive a relaxation equation for the bulk viscous pressure, we use
the same technique applied to the shear stress tensor, \ie, we
replace $\theta\simeq -\zeta^{-1}{\Pi}$ in the first term of
Eq.~\eqref{eq:bulk_av_2_final}. We then obtain ($\dot{\Pi}\equiv D\Pi$)
\bea\label{eq:bulk_relax}
\zeta\tau_\Pi D\theta &=& 
-\tau_\Pi \dot{\Pi} +\tau_\Pi{\Pi}\zeta^{-1}D\zeta
\nonumber\\
&=&-\tau_\Pi \dot{\Pi}+\tau_\Pi \beta \zeta^{-1}
\bigg(\gamma\frac{\partial\zeta}{\partial\beta} 
-\sum\limits_a \delta_a\frac{\partial\zeta}
{\partial\alpha_a}\bigg)\theta \Pi,
\eea
where we used Eqs.~\eqref{eq:D_beta1} and \eqref{eq:D_alpha1}.
Combining Eqs.~\eqref{eq:bulk_sum}, \eqref{eq:bulk_av_2_final} 
and \eqref{eq:bulk_relax}, and defining
\bea\label{eq:varsigma}
\varsigma &=& \lambda_\Pi +\zeta^*  +
\psi_{\epsilon\epsilon}\zeta_\epsilon^2 +
2\zeta_\epsilon \sum\limits_a\psi_{\epsilon a} \zeta_{a} +
\sum\limits_{ab}\psi_{ab} \zeta_{a} \zeta_{b},\\
\label{eq:tilde_lambda_Pi}
\tilde{\lambda}_\Pi &=& \tau_{\Pi}\beta \zeta^{-1}
\bigg(\gamma\frac{\partial\zeta}{\partial\beta} 
-\sum\limits_a \delta_a\frac{\partial\zeta}
{\partial\alpha_a}\bigg),
\eea
 we obtain finally
\bea\label{eq:bulk_final_relax}
\tau_\Pi \dot{\Pi}+\Pi &=&
-\zeta\theta +  \tilde{\lambda}_\Pi \theta \Pi 
+\zeta_\beta (\sigma_{\mu\nu}\pi^{\mu\nu}-\theta \Pi) -\tilde{\zeta}_\beta \partial_\mu q^{\mu}+
\varsigma\theta^2  \nonumber\\
&&-\lambda_{\Pi\pi} \sigma_{\mu\nu}\sigma^{\mu\nu}+\sum_a \bar{\zeta}_a
\dot{u}_{\rho} \nabla^\rho\alpha_a  +
\sum\limits_a \zeta_{\alpha_a} \partial_\mu {\mathscr J}^{\mu}_a  \nonumber\\
&& +T\sum\limits_{ab}\zeta^{ab}_\Pi
\nabla^\mu \alpha_a\nabla_\mu\alpha_b 
+q^\mu \Big[\zeta_\beta \dot{u}_\mu +\sum\limits_a \zeta_{\alpha_a} 
\nabla_\mu (n_ah^{-1})\Big],
\eea
where the second term on the middle line is new.

\subsection{Second-order corrections to the diffusion currents}
\label{sec:2nd_diff}

Using Eqs.~\eqref{eq:C_deriv1} and \eqref{eq:stat_average_21} we obtain
\bea\label{eq:current_av_non_local1}
\langle \hat{\mathscr J}_{c\mu}(x)\rangle_2^1
&=& -\Delta_{\mu\rho}(x)\sum\limits_a
\Big[\partial_{\tau}(\nabla^\rho\alpha_a)-
\beta\theta (\partial_{\tau}u^\rho) \delta_a \Big]_x \nonumber\\
&&\times \frac{1}{3} \int\! d^4x_1\Big(\hat{\mathscr J}_{c\lambda}(x),
 \hat{\mathscr{J}}^{\lambda}_a(x_1)_x \Big)(x_1-x)^\tau
  \nonumber\\
&& +\Delta_{\mu\rho}(x)\Big[2\beta \sigma^{\rho\sigma}
(\partial_{\tau} u_\sigma)
+y\beta \theta (\partial_{\tau} u^\rho) -\sum_a \partial_{\tau}
(n_ah^{-1})
\nabla^\rho\alpha_a\Big]_x \nonumber\\
&& \times \frac{1}{3}
\int\! d^4x_1\Big(\hat{\mathscr J}_{c\lambda}(x),
\hat{q}^{\lambda}(x_1)_x \Big)(x_1-x)^\tau,
\eea
where
\bea
&& y=\frac{2}{3}-2\gamma -\sum\limits_a \delta_a   n_a h^{-1},\\
&& \frac{1}{3} \int\! d^4x_1\Big(\hat{\mathscr J}_{c\lambda}(x),
 \hat{\mathscr{J}}^{\lambda}_a(x_1)_x \Big)(x_1-x)^\tau = -u^\tau
 \tilde{\chi}_{ca},\\
 &&\frac{1}{3}
\int\! d^4x_1\Big(\hat{\mathscr J}_{c\lambda}(x),
\hat{q}^{\lambda}(x_1)_x \Big)(x_1-x)^\tau = -u^\tau \tilde{\chi}_{c},
\eea
and we used Eq.~\eqref{eq:corr1_current} 
and an analogous relation  
\bea\label{eq:corr2_current}
\Big(\hat{\mathscr J}_{c\mu}(x),
\hat{q}_{\rho}(x_1)_x\Big) =\frac{1}{3}
\Delta_{\mu\rho}(x) \Big(\hat{\mathscr J}_{c\lambda}(x),
\hat{q}^{\lambda}(x_1)_x\Big).
\eea
Next, using also Eqs.~\eqref{eq:vector_I0_rest}--\eqref{eq:vec_gen} we obtain 
\bea\label{eq:current_av_non_local2}
\langle \hat{\mathscr J}_{c\mu}\rangle_2^1
&=& \sum\limits_a\tilde{\chi}_{ca}
\Delta_{\mu\rho} D(\nabla^\rho\alpha_a)  
-\tilde{\chi}_{c}\sum_a D(n_ah^{-1})\nabla_\mu\alpha_a\nonumber\\
&& 
-\beta\theta  \dot{u}_\mu\sum\limits_a\delta_a
\tilde{\chi}_{ca} +\tilde{\chi}_{c}\beta
(2\sigma_{\mu\nu} \dot{u}^\nu +y \theta \dot{u}_\mu),
\eea
where
\bea\label{eq:chi_ab_tilde}
\tilde{\chi}_{ca} &=& i\frac{d}{d\omega} 
{\chi}_{ca}(\omega)\bigg\vert_{\omega=0}=
\frac{T}{6}\frac{d^2}{d\omega^2} {\rm Re}
G^R_{\mathscr{\hat{J}}_{c}^{\lambda}
\mathscr{\hat{J}}_{a\lambda}}(\omega)
\bigg\vert_{\omega=0},\\
\label{eq:chi_c_tilde}
\tilde{\chi}_{c} &=& i\frac{d}{d\omega} 
{\chi}_{c}(\omega)\bigg\vert_{\omega=0}=
\frac{T}{6} \frac{d^2}{d\omega^2} {\rm Re}
G^R_{\hat{\mathscr J}_{c}^{\lambda}
{\hat{q}}_{\lambda}}(\omega)\bigg\vert_{\omega=0},\\
\label{eq:chi_c}
\chi_{c} &=& -\frac{1}{3} \int d^4x_1
\Big (\hat{\mathscr J}_c^{\lambda}(x),
\hat{q}_{\lambda}(x_1)_x\Big) =\frac{T}{3}
\frac{d}{d\omega} {\rm Im}G^R_{\hat{\mathscr{J}}_c^{\lambda}\hat{q}_{\lambda}}(\omega)\bigg\vert_{\omega=0}.
\eea

Further, one can utilize Eq.~\eqref{eq:ideal_hydro2} to 
write  $D (n_ah^{-1})=-n_ah^{-2}Dp $. From 
Eqs.~\eqref{eq:ideal_hydro2} and \eqref{eq:gamma_delta_a} we find
\bea\label{eq:Dp}
Dp=\gamma D\epsilon+\sum\limits_a \delta_a Dn_a
=-\Big(\gamma h+\sum\limits_a \delta_a n_a\Big)\theta.
\eea
Substituting these results into Eqs.~\eqref{eq:current_av_non_local2} we obtain 
\bea\label{eq:current_av_21}
\langle \hat{\mathscr J}_{c\mu}\rangle_2^1 
=
\sum\limits_a \tilde{\chi}_{ca}\Delta_{\mu\beta}
D (\nabla^\beta\alpha_a)
-
\tilde{\chi}_{c} h^{-2} \Big(\gamma h+\sum\limits_b 
\delta_b n_b\Big)\theta \sum\limits_a 
n_a \nabla_\mu\alpha_a\nonumber\\
-\beta\theta  \dot{u}_\mu\sum\limits_a
\delta_a\tilde{\chi}_{ca} + \tilde{\chi}_{c}\beta
(2\sigma_{\mu\nu}\dot{u}^\nu+y\theta\dot{u}_\mu).
\eea
The second line in this expression collects the new terms, among which
the terms $\propto \theta \dot{u}_\mu$ are responsible for the
non-local mixing of charge diffusion currents with the bulk viscous
pressure, and the term $\propto \sigma_{\mu\nu} \dot{u}^\nu$
corresponds to the non-local mixing of charge diffusion currents with
the shear stresses.

The averages $\langle \hat{\mathscr J}_{c\mu}\rangle_2^2$
and $\langle \hat{\mathscr J}_{c\mu}\rangle_2^3$
are given by~\cite{Harutyunyan_hydro2022}
\bea\label{eq:current_av_22}
\langle \hat{\mathscr J}_{c\mu}\rangle_2^2 &=&
 \chi_{c} \beta h^{-1}(-\nabla_\mu \Pi
+\Pi \dot{u}_\mu + \dot{q}_{\mu}
 + q^{\nu}\partial_\nu u_{\mu}+ q_{\mu}\theta 
 +\Delta_{\mu\sigma}\partial_\nu \pi^{\nu\sigma}),\\
\label{eq:current_av_23}
\langle \hat{\mathscr J}_{c\mu}\rangle_2^3 &=&
\sum\limits_a\Big( \zeta_{{\!\mathscr J}}^{ca} 
\theta\, \nabla_\mu\alpha_a-\lambda_{{\!\mathscr J}}^{ca} 
\sigma_{\mu\nu} \nabla^\nu\alpha_a \Big),
\eea
where we defined new coefficients via
\bea\label{eq:zeta_3_ca}
\zeta_{{\!\mathscr J}}^{ca} &=& 
\frac{2\beta }{3}\int d^4x_1d^4x_2
\Big(\hat{\mathscr J}_{c\gamma}(x),
\hat{\mathscr J}_{a}^\gamma (x_1)_x,
\hat{p}^*(x_2)_x\Big),\\
\label{eq:lambda_3_ca}
\lambda_{{\!\mathscr J}}^{ca} &=& 
\frac{2\beta}{5}\int d^4x_1d^4x_2
\Big(\hat{\mathscr J}_{c}^{\gamma}(x),
\hat{\mathscr J}_{a}^{\delta}(x_1)_x,
\hat{\pi}_{\gamma\delta}(x_2)_x\Big).
\eea

\subsubsection{Final equation for the diffusion currents}
\label{sec:final_diff}
 
Combining Eqs.~\eqref{eq:stat_average_2nd}, \eqref{eq:stat_average_2_new},
\eqref{eq:shear_bulk_diff}, \eqref{eq:current_av_21}, 
\eqref{eq:current_av_22} and \eqref{eq:current_av_23} we obtain the 
diffusion currents up to the second order in hydrodynamic gradients
\bea\label{eq:current_total_final}
{\mathscr J}_{c\mu}(x)
&=&\sum\limits_b\chi_{cb}\nabla_\mu \alpha_b +
\sum\limits_a \tilde{\chi}_{ca}\Delta_{\mu\beta}
D (\nabla^\beta\alpha_a)-\tilde{\chi}_{c} h^{-2} 
\Big(\gamma h+\sum\limits_b \delta_b n_b\Big) \nonumber\\
&& \times \theta\sum\limits_a 
n_a \nabla_\mu\alpha_a
-\beta\theta  \dot{u}_\mu\sum\limits_a
\delta_a\tilde{\chi}_{ca} +\tilde{\chi}_{c}\beta
(2\sigma_{\mu\nu} \dot{u}^\nu +y \theta \dot{u}_\mu)\nonumber\\
&&
+\sum\limits_a\Big( \zeta_{{\!\mathscr J}}^{ca} 
\theta\, \nabla_\mu\alpha_a -\lambda_{{\!\mathscr J}}^{ca}
\sigma_{\mu\nu} \nabla^\nu\alpha_a \Big)\nonumber\\
&&
+\chi_{c}\beta h^{-1}(-\nabla_\mu \Pi
+\Pi \dot{u}_\mu + \dot{q}_{\mu}
+ q^{\nu}\partial_\nu u_{\mu} + q_{\mu}\theta 
+\Delta_{\mu\sigma}\partial_\nu\pi^{\nu\sigma}).
\eea
To obtain relaxation equations for the diffusion currents we 
modify the second term in Eq.~\eqref{eq:current_total_final}
by employing the third equation in Eq.~\eqref{eq:shear_bulk_diff} 
in the form 
\bea\label{eq:current_invert}
\nabla^\beta \alpha_a 
=\sum\limits_b(\chi^{-1})_{ab}\mathscr{J}_b^\beta.
\eea 
Then, utilizing in addition Eqs.~\eqref{eq:D_beta1} and \eqref{eq:D_alpha1} 
at the leading order, we obtain
\bea\label{eq:diff_relax1}
&& \sum\limits_a \tilde{\chi}_{ca}\Delta_{\mu\beta}
D (\nabla^\beta\alpha_a)
=-\sum\limits_b \tau_{\!\mathscr J}^{cb}
\mathscr{\dot{J}}_{b\mu}\nonumber\\
&&+ \beta\theta \sum\limits_{ab} \tilde{\chi}_{ca}
\left[\gamma\frac{\partial(\chi^{-1})_{ab}}{\partial \beta}
-\sum\limits_d \delta_d\frac{\partial(\chi^{-1})_{ab}}
{\partial \alpha_d}\right]\!\mathscr{J}_{ b\mu}, 
\eea 
where $\mathscr{\dot{J}}_{a\mu}=\Delta_{\mu\nu} D\!\mathscr{J}^\nu_a$, 
and we defined a matrix of relaxation times 
\bea\label{eq:diff_ab_relax}
\tau_{\!\mathscr J}^{cb} = -
(\tilde{\chi}\chi^{-1})_{cb}=
-\sum\limits_a \tilde{\chi}_{ca}(\chi^{-1})_{ab}.
\eea
Introducing also the coefficients
\bea\label{eq:lambda_tilde}
\tilde{\lambda}^{cb}_{\!\mathscr J} &=& \beta
\sum\limits_{a} \tilde{\chi}_{ca}
\left[\gamma\frac{\partial(\chi^{-1})_{ab}}{\partial \beta}
-\sum\limits_d \delta_d\frac{\partial(\chi^{-1})_{ab}}
{\partial \alpha_d}\right],\\
\label{eq:chi_star}
\chi^*_{cb} &=& 
\zeta_{{\!\mathscr J}}^{cb}-\tilde{\chi}_{c}n_b h^{-2} 
\Big(\gamma h+\sum\limits_d \delta_d n_d\Big),
\eea
we obtain from Eqs.~\eqref{eq:current_total_final} and \eqref{eq:diff_relax1}
\bea\label{eq:current_final_relax}
\sum\limits_b \tau_{\!\mathscr J}^{ab}\!\mathscr{\dot{J}}_{b\mu}
+{\mathscr J}_{a\mu}
=\sum\limits_b\Big[\chi_{ab}\nabla_\mu \alpha_b 
+  \tilde{\lambda}^{ab}_{\!\mathscr J} 
\theta\!\mathscr{J}_{ b\mu} + \chi^*_{ab} 
\theta\nabla_\mu\alpha_b -
{\lambda}^{ab}_{\!\mathscr J} 
 \sigma_{\mu\nu} \nabla^\nu\alpha_b 
 -\beta\theta  \dot{u}_\mu
\tilde{\chi}_{ab} \delta_b \Big]\nonumber\\
+ 
\chi_{a}\beta h^{-1}\big(-\nabla_\mu \Pi
+\Pi \dot {u}_\mu + \dot{q}_{\mu}
 + q^{\nu}\partial_\nu u_{\mu}+ q_{\mu}\theta 
 +\Delta_{\mu\sigma}\partial_\nu \pi^{\nu\sigma}\big)
 +\tilde{\chi}_{a}\beta(2\sigma_{\mu\nu} \dot{u}^\nu +y \theta \dot{u}_\mu).\quad
\eea
If there is only one sort of conserved charge, then 
Eq.~\eqref{eq:current_final_relax} simplifies to
\bea\label{eq:current_final_relax1}
\tau_{\!\mathscr J}\!\mathscr{\dot{J}}_{\mu}+
{\mathscr J}_{\mu} = \chi \nabla_\mu \alpha 
+  \tilde{\lambda}_{\!\mathscr J} 
\theta\!\mathscr{J}_{\mu} + \chi^* 
\theta\nabla_\mu\alpha -{\lambda}_{\!\mathscr J} 
 \sigma_{\mu\nu} \nabla^\nu\alpha 
 -\beta\theta \tilde{\chi} \delta\dot{u}_\mu\nonumber\\
+
\chi'\beta h^{-1}\big(-\nabla_\mu \Pi
+\Pi \dot {u}_\mu + \dot{q}_{\mu}
 + q^{\nu}\partial_\nu u_{\mu}+ q_{\mu}\theta 
 +\Delta_{\mu\sigma}\partial_\nu \pi^{\nu\sigma}\big)
 +\tilde{\chi}'\beta(2\sigma_{\mu\nu} \dot{u}^\nu +y \theta \dot{u}_\mu),
\eea
where the current relaxation time is given by 
[see Eqs.~\eqref{eq:chi_ab_tilde} and \eqref{eq:diff_ab_relax}] 
\bea\label{eq:diff_current_relax}
\chi \tau_{\!\mathscr J} = 
 -i\frac{d}{d\omega}\chi(\omega)\bigg\vert_{\omega=0}
=-\frac{T}{6}\frac{d^2}{d\omega^2} {\rm Re}G^R_{\hat{{\mathscr J}_{\mu}}
\hat{\mathscr J}^{\mu}}(\omega)\bigg\vert_{\omega=0}, 
\eea
and 
\bea\label{eq:lambda_tilde1}
\tilde{\lambda}_{\!\mathscr J} &=& \tau_{\!\mathscr J} 
\beta\chi^{-1}\left(\gamma\frac{\partial\chi}{\partial \beta}
-\delta\frac{\partial\chi}{\partial\alpha}\right),\\
\label{eq:chi_star1}
\chi^* &=& \zeta_{{\!\mathscr J}}-\tilde{\chi}'
n h^{-2} (\gamma h+\delta n).
\eea
Note that the diffusion coefficient and the thermal 
conductivity are related via 
$\kappa=\left(\frac{h}{nT}\right)^2\chi$.
The frequency-dependent coefficient $\chi$ in 
Eqs.~\eqref{eq:diff_current_relax} is defined 
according to the formula~\eqref{eq:I_XY} in 
\ref{app:Green_func} with the relevant choice of operators.


\subsection{Second-order corrections to the energy flux}
\label{sec:2nd_diff}

For the sake of completeness, we derive an equation also for 
the energy flux $q_{\mu}$. The derivation is quite analogous to
that for the diffusion currents. Using Eqs.~\eqref{eq:stat_average_21},
\eqref{eq:C_deriv1} and \eqref{eq:vector_I0_rest}--\eqref{eq:vec_gen}
and exploiting Curie's theorem again we can obtain
\bea\label{eq:flux_av_non_local2}
\langle \hat{q}_{\mu}(x)\rangle_2^1
&=& \sum\limits_a\tilde{\chi}_{a}
\Delta_{\mu\rho} D(\nabla^\rho\alpha_a)  
-\tilde{\chi}_{q}\sum_a D(n_ah^{-1})\nabla_\mu\alpha_a\nonumber\\
&& -\beta\theta  \dot{u}_\mu\sum\limits_a
\delta_a\tilde{\chi}_{a} +\tilde{\chi}_{q}\beta
(2\sigma_{\mu\nu} \dot{u}^\nu +y \theta \dot{u}_\mu),
\eea
where we employed Eq.~\eqref{eq:corr2_current} 
and an analogous relation 
\bea\label{eq:corr3_current}
\Big(\hat{q}_{\mu}(x),\hat{q}_{\rho}(x_1)_x\Big) =
\frac{1}{3}\Delta_{\mu\rho}(x) \Big(\hat{q}_{\lambda}(x),
\hat{q}^{\lambda}(x_1)_x\Big),
\eea
and defined the transport coefficients
\bea\label{eq:chi_q}
\tilde{\chi}_{q} &=& i\frac{d}{d\omega} 
{\chi}_{q}(\omega)\bigg\vert_{\omega=0}=
\frac{T}{6} \frac{d^2}{d\omega^2} {\rm Re}
G^R_{\hat{q}^{\lambda}
{\hat{q}}_{\lambda}}(\omega)\bigg\vert_{\omega=0},\\
\label{eq:chi_q_tilde}
\chi_{q} &=& -\frac{1}{3} \int\! d^4x_1
\Big (\hat{q}^{\lambda}(x),
\hat{q}_{\lambda}(x_1)_x\Big) =\frac{T}{3}
\frac{d}{d\omega} {\rm Im}G^R_{\hat{q}^{\lambda}
\hat{q}_{\lambda}}(\omega)\bigg\vert_{\omega=0}.
\eea
Substituting Eq.~\eqref{eq:Dp} and the expression above it
in Eq.~\eqref{eq:flux_av_non_local2} we find
\bea\label{eq:flux_av_21}
\langle \hat{q}_{\mu}\rangle_2^1 
=
\sum\limits_a \tilde{\chi}_{a}\Delta_{\mu\beta}
D (\nabla^\beta\alpha_a)
-
\tilde{\chi}_{q} h^{-2} \Big(\gamma h+\sum\limits_b 
\delta_b n_b\Big)\theta \sum\limits_a 
n_a \nabla_\mu\alpha_a\nonumber\\
-\beta\theta  \dot{u}_\mu\sum\limits_a
\delta_a\tilde{\chi}_{a} + \tilde{\chi}_{q}\beta
(2\sigma_{\mu\nu} \dot{u}^\nu +y \theta \dot{u}_\mu).
\eea

The averages $\langle \hat{q}_{\mu}\rangle_2^2$ and 
$\langle \hat{q}_{\mu}\rangle_2^3$ can be computed 
according to Eqs.~\eqref{eq:stat_average_22} and 
\eqref{eq:stat_average_23} 
\bea\label{eq:flux_av_22}
\langle \hat{q}_{\mu}\rangle_2^2 &=&
 \chi_{q} \beta h^{-1}(-\nabla_\mu \Pi
+\Pi \dot{u}_\mu + \dot{q}_{\mu}
 + q^{\nu}\partial_\nu u_{\mu}+ q_{\mu}\theta 
 +\Delta_{\mu\sigma}\partial_\nu \pi^{\nu\sigma}),\\
\label{eq:flux_av_23}
\langle \hat{q}_{\mu}\rangle_2^3 &=&
\sum\limits_a\Big( \zeta_{q}^{a} \theta\, \nabla_\mu\alpha_a
-\lambda_{q}^{a} \sigma_{\mu\nu} \nabla^\nu\alpha_a \Big),
\eea
where we defined new coefficients via
\bea\label{eq:zeta_q_a}
\zeta_{q}^{a} &=& \frac{2\beta }{3}\!\int\! 
d^4x_1d^4x_2 \Big(\hat{q}_{\gamma}(x),
\hat{\mathscr J}_{a}^\gamma (x_1)_x,
\hat{p}^*(x_2)_x\Big),\\
\label{eq:lambda_q_a}
\lambda_{q}^{a} &=& \frac{2\beta}{5}\!\int\! 
d^4x_1d^4x_2 \Big(\hat{q}^{\gamma}(x),
\hat{\mathscr J}_{a}^{\delta}(x_1)_x,
\hat{\pi}_{\gamma\delta}(x_2)_x\Big).
\eea

Note that from Eqs.~\eqref{eq:op_C1}, \eqref{eq:stat_average_C1}
and \eqref{eq:chi_c} the first-order correction to the 
energy flux is given by
\bea \label{eq:flux1}
\langle\hat{q}^\mu\rangle_1
= \sum\limits_b\chi_{b}\nabla^\mu \alpha_b.
\eea

\subsubsection{Final expression for the energy flux}
\label{sec:final_diff}

Combining Eqs.~\eqref{eq:stat_average_2nd}, 
\eqref{eq:stat_average_2_new},
\eqref{eq:flux1}, \eqref{eq:flux_av_21}, 
\eqref{eq:flux_av_22} and \eqref{eq:flux_av_23} we obtain the 
energy flux up to the second order in hydrodynamic gradients
\bea\label{eq:flux_total_final}
{q}_{\mu}(x)
&=&\sum\limits_b\chi_{b}\nabla_\mu \alpha_b +
\sum\limits_a \tilde{\chi}_{a}\Delta_{\mu\beta}
D (\nabla^\beta\alpha_a)-\tilde{\chi}_{q} h^{-2} 
\Big(\gamma h+\sum\limits_b \delta_b n_b\Big) \nonumber\\
& \times &\theta\sum\limits_a 
n_a \nabla_\mu\alpha_a
-\beta\theta  \dot{u}_\mu\sum\limits_a
\delta_a\tilde{\chi}_{a} +\tilde{\chi}_{q}\beta
(2\sigma_{\mu\nu} \dot{u}^\nu +y \theta \dot{u}_\mu)\nonumber\\
&+&
\sum\limits_a\Big( \zeta_{a}^{a} 
\theta\, \nabla_\mu\alpha_a
-\lambda_{q}^{a} \sigma_{\mu\nu} 
\nabla^\nu\alpha_a \Big)\nonumber\\
&+&
\chi_{q}\beta h^{-1}(-\nabla_\mu \Pi
+\Pi \dot{u}_\mu + \dot{q}_{\mu}
 + q^{\nu}\partial_\nu u_{\mu}+ q_{\mu}\theta 
 +\Delta_{\mu\sigma}\partial_\nu \pi^{\nu\sigma}).
\eea
We next substitute Eq.~\eqref{eq:current_invert} in the 
second term in Eq.~\eqref{eq:flux_total_final} 
and define the coefficients 
\bea\label{eq:lambda_q_tilde}
\tilde{\lambda}^{b}_{q} &=& \beta
\sum\limits_{a} \tilde{\chi}_{a}
\left[\gamma\frac{\partial(\chi^{-1})_{ab}}{\partial \beta}
-\sum\limits_d \delta_d\frac{\partial(\chi^{-1})_{ab}}
{\partial \alpha_d}\right],\\
\label{eq:flux_ab_relax}
\tau_{q}^{b} &=& 
-n_bh^{-1}\sum\limits_a 
\tilde{\chi}_{a}(\chi^{-1})_{ab},
\eea
to obtain
\bea\label{eq:flux_relax1}
\sum\limits_a \tilde{\chi}_{a}\Delta_{\mu\beta}
D (\nabla^\beta\alpha_a)
&=& -\sum\limits_{b} 
hn_b^{-1}\tau_{q}^{b}
\mathscr{\dot{J}}_{b\mu}+ \theta \sum\limits_{b} 
\tilde{\lambda}^{b}_{q} \mathscr{J}_{ b\mu}.
\eea 
Introducing also the coefficients
\bea\label{eq:chi_q_star}
\chi^*_{b} &=& 
\zeta_{q}^{b}-\tilde{\chi}_{q}n_b h^{-2} 
\Big(\gamma h+\sum\limits_d \delta_d n_d\Big),
\eea
we obtain from Eqs.~\eqref{eq:flux_total_final} 
and \eqref{eq:flux_relax1}
\bea\label{eq:flux_final_relax}
{q}_{\mu} =\sum\limits_b\Big[\chi_{b}\nabla_\mu \alpha_b 
-hn_b^{-1}\tau_{q}^{b}\mathscr{\dot{J}}_{b\mu}
+  \tilde{\lambda}^{b}_{q} 
\theta\! \mathscr{J}_{ b\mu} + \chi^*_{b} 
\theta\,\nabla_\mu\alpha_b -
{\lambda}^{b}_{q} \sigma_{\mu\nu} \nabla^\nu\alpha_b 
 - \beta\theta  \dot{u}_\mu
\tilde{\chi}_{b} \delta_b \Big]\nonumber\\
+ 
\chi_{q}\beta h^{-1}\big(-\nabla_\mu \Pi
+\Pi \dot {u}_\mu + \dot{q}_{\mu}
 + q^{\nu}\partial_\nu u_{\mu}+ q_{\mu}\theta 
 +\Delta_{\mu\sigma}\partial_\nu \pi^{\nu\sigma}\big)
 + \tilde{\chi}_{q}\beta
(2\sigma_{\mu\nu} \dot{u}^\nu +y \theta \dot{u}_\mu).
\eea
If we have only one sort of conserved charge, then 
Eq.~\eqref{eq:flux_final_relax} simplifies to
\bea\label{eq:flux_final_relax1}
{q}_{\mu} &=& \chi'\nabla_\mu \alpha
-hn^{-1}\tau_{q}\mathscr{\dot{J}}_{\mu}
+  \tilde{\lambda}'_{q} 
\theta\! \mathscr{J}_{\mu} + 
\chi^* \theta\,\nabla_\mu\alpha -
{\lambda}'_{q} \sigma_{\mu\nu} \nabla^\nu\alpha
 -\beta\theta  \dot{u}_\mu
\tilde{\chi} \delta 
+\chi_{q}\beta h^{-1}\nonumber\\
&& \times \big(-\nabla_\mu \Pi
+\Pi \dot {u}_\mu + \dot{q}_{\mu}
 + q^{\nu}\partial_\nu u_{\mu}+ q_{\mu}\theta 
 +\Delta_{\mu\sigma}\partial_\nu \pi^{\nu\sigma}\big)
 +\tilde{\chi}_{q}\beta
(2\sigma_{\mu\nu} \dot{u}^\nu +y \theta \dot{u}_\mu).\quad
\eea

\section{Discussion and conclusions}
\label{sec:hydro_discuss}

\subsection{General structure of the transport equations}
\label{sec:discuss_eqs}

Here we write down the complete set of second-order transport
equations for the shear stress tensor, the bulk viscous pressure, the
diffusion fluxes, and the energy flux, respectively [see
Eqs.~\eqref{eq:shear_final_relax}, \eqref{eq:bulk_final_relax},
\eqref{eq:current_final_relax} and \eqref{eq:flux_final_relax}]
\bea\label{eq:shear_final}
\tau_\pi \dot{\pi}_{\mu\nu}+ {\pi}_{\mu\nu}
&=& 2\eta \sigma_{\mu\nu}
+\tilde{\lambda}_\pi\theta\pi_{\mu\nu}
- 2\eta\tau_\pi Th^{-1}\sum_a n_a
\dot{u}_{<\mu} \nabla_{\nu >}\alpha_a\nonumber\\ 
&& + \lambda \theta {\sigma}_{\mu\nu} + 
 \lambda_\pi \sigma_{\rho<\mu}\sigma_{\nu>}^{\rho} 
 +\sum\limits_{ab}\lambda_{\pi{\!\mathscr J}}^{ab}
 \nabla_{<\mu}\alpha_a\nabla_{\nu>}\alpha_b,\\
\label{eq:bulk_final}
\tau_\Pi \dot{\Pi}+\Pi &=&
-\zeta\theta +  \tilde{\lambda}_\Pi \theta \Pi 
+\varsigma\theta^2 +\zeta_\beta (\sigma_{\mu\nu}
\pi^{\mu\nu}-\theta \Pi)  -\lambda_{\Pi\pi} 
\sigma_{\mu\nu}\sigma^{\mu\nu}\nonumber\\
&& + \sum\limits_a \zeta_{\alpha_a} 
\partial_\mu {\mathscr J}^{\mu}_a 
-\tilde{\zeta}_\beta \partial_\mu q^{\mu}
+\sum_a \bar{\zeta}_a
\dot{u}_{\mu} \nabla^\mu\alpha_a \nonumber\\
 &&  +q^\mu \Big[\zeta_\beta \dot{u}_\mu +
 \sum\limits_a \zeta_{\alpha_a} \nabla_\mu (n_ah^{-1})\Big]
 +T\sum\limits_{ab}\zeta^{ab}_\Pi
\nabla^\mu \alpha_a\nabla_\mu\alpha_b,\qquad\qquad\\
\label{eq:current_final}
\sum\limits_b\tau_{\!\mathscr J}^{ab}
\mathscr{\dot{J}}_{b\mu}+{\mathscr J}_{a\mu}
&=&\sum\limits_b\Big[\chi_{ab}\nabla_\mu \alpha_b 
+  \tilde{\lambda}^{ab}_{\!\mathscr J} 
\theta\!\mathscr{J}_{ b\mu} + \chi^*_{ab}
\theta\nabla_\mu\alpha_b -
{\lambda}^{ab}_{\!\mathscr J} 
 \sigma_{\mu\nu} \nabla^\nu\alpha_b \nonumber\\
&& -\beta\theta  \dot{u}_\mu
\tilde{\chi}_{ab} \delta_b\Big]
+\tilde{\chi}_{a}\beta
(2\sigma_{\mu\nu} \dot{u}^\nu + y\theta \dot{u}_\mu)\nonumber\\
&& +
\chi_{a}\beta h^{-1}\big(\Pi \dot {u}_\mu-\nabla_\mu \Pi
 + \dot{q}_{\mu} + q^{\nu}\partial_\nu u_{\mu}+ q_{\mu}\theta 
 +\Delta_{\mu\sigma}\partial_\nu \pi^{\nu\sigma}\big),\hspace{2cm}\\
\label{eq:flux_final}
{q}_{\mu} &=& \sum\limits_b\Big[\chi_{b}\nabla_\mu \alpha_b 
-hn_b^{-1}\tau_{q}^{b}\mathscr{\dot{J}}_{b\mu}
+  \tilde{\lambda}^{b}_{q} 
\theta\! \mathscr{J}_{ b\mu} + \chi^*_{b} 
\theta\,\nabla_\mu\alpha_b \nonumber\\
&& -{\lambda}^{b}_{q} \sigma_{\mu\nu} \nabla^\nu\alpha_b -\beta\theta  \dot{u}_\mu
\tilde{\chi}_{b} \delta_b \Big]
 +\tilde{\chi}_{q}\beta
(2\sigma_{\mu\nu} \dot{u}^\nu +y \theta \dot{u}_\mu)\nonumber\\
&& + 
\chi_{q}\beta h^{-1}\big(-\nabla_\mu \Pi
+\Pi \dot {u}_\mu + \dot{q}_{\mu}
 + q^{\nu}\partial_\nu u_{\mu}+ q_{\mu}\theta 
 +\Delta_{\mu\sigma}\partial_\nu \pi^{\nu\sigma}\big),
\eea
where the dot denotes the comoving derivative
\bea\label{eq:corrent_dot}
\dot{\Pi}=D\Pi,\quad
\dot{\pi}_{\mu\nu}=\Delta_{\mu\nu\rho\sigma}
D{\pi}^{\rho\sigma},\quad
\dot{u}_\mu = Du_\mu,\quad
\dot{q}_\mu = \Delta_{\mu\nu} D q^{\nu},\quad
\mathscr{\dot{J}}_{a\mu}=\Delta_{\mu\nu} D\!\mathscr{J}^\nu_a.
\eea

The initial terms on the right-hand sides of
Eqs.~\eqref{eq:shear_final}--\eqref{eq:current_final} represent the
Navier--Stokes' contributions to the dissipative quantities. These
first-order coefficients include shear viscosity $\eta$, bulk
viscosity $\zeta$ and the diffusion coefficient matrix $\chi_{ab}$,
$\chi_{b}$. These coefficients are derived using two-point retarded
correlation functions via the Kubo
formulas~\eqref{eq:shear_bulk_mod}--\eqref{eq:green_func}, and
\eqref{eq:chi_c}.  The first terms on the left-hand sides of
Eqs.~\eqref{eq:shear_final}--\eqref{eq:current_final} capture the
relaxation of the dissipative fluxes towards their Navier-Stokes
values, characterized by specific relaxation time scales $\tau_\pi$,
$\tau_\Pi$ and $\tau_{\!\mathscr J}^{ab}$. These timescales are
related to the relevant first-order transport coefficients by
Eqs.~\eqref{eq:eta_relax}, \eqref{eq:zeta_relax} and
\eqref{eq:diff_ab_relax}.  The memory effects within the
non-equilibrium statistical operator give rise to these relaxation
expressions. When the system retains information about its past states
(finite memory), this manifests as dispersion (\ie,
frequency-dependence) of the first-order transport. The second terms on
the right-hand sides of
Eqs.~\eqref{eq:shear_final}--\eqref{eq:current_final} appear when the
first-order transport coefficients vary in space and/or time, as they
generally depend on temperature and chemical potentials, which
themselves vary across the system and in time.  The coefficients
$\tilde{\lambda}_\pi$, $\tilde{\lambda}_\Pi$ and
$\tilde{\lambda}^{ab}_{\!\mathscr J}$ which stand in front of these
terms are, therefore, expressed in terms of the derivatives of the
corresponding first-order transport coefficients with respect to the
temperature and the chemical potentials by
Eqs.~\eqref{eq:tilde_lambda_pi}, \eqref{eq:tilde_lambda_Pi} and
\eqref{eq:lambda_tilde}.

In Eqs.~\eqref{eq:shear_final}--\eqref{eq:current_final} three
distinct types of second-order terms emerge that do not represent
relaxation processes: (i) terms combining thermodynamic forces with
dissipative fluxes (such as the term proportional to $\theta \Pi$) in
Eq.~\eqref{eq:bulk_final}, (ii) terms involving spatial derivatives of
dissipative fluxes (like $\partial_\mu {\mathscr J}^{\mu}_a$), and
(iii) non-linear, quadratic terms in thermodynamic forces (for
example, $\theta\sigma_{\mu\nu}$ ).

The terms of the types (i) originate either
from the non-local corrections~\eqref{eq:stat_average_21} [second
terms on the right-hand sides of
Eqs.~\eqref{eq:shear_final}--\eqref{eq:current_final}], or from the
corrections which include the extended thermodynamic force
$\hat{C}_2$~\eqref{eq:stat_average_22}.  The corrections of the type
(ii) arise purely from the operator $\hat{C}_2$.  The transport
coefficients in terms of type (i) and (ii) are related to
two-point correlation functions by Eqs.~\eqref{eq:zeta_beta}--\eqref{eq:tilde_zeta_beta} and \eqref{eq:chi_c}.
The corrections of the type (iii) arise from two sources. Firstly,
such terms arise from the quadratic term of the second-order expansion
of the statistical operator, which corresponds to the statistical
average given by Eq.~\eqref{eq:stat_average_23}. These terms contain
all possible combinations that are quadratic in the first-order
thermodynamic forces $\sigma_{\mu\nu}$, $\theta$, and
$\nabla_\mu\alpha_a$. For example, the relevant corrections for the
shear stress tensor contain the three combinations
$\theta {\sigma}_{\mu\nu}$, $\sigma_{\rho<\mu}\sigma_{\nu>}^{\rho}$
and $\nabla_{<\mu}\alpha_a\nabla_{\nu>}\alpha_b$.  The transport
coefficients related to these terms are characterized by three-point
correlation functions, which reveal the intricate, nonlinear
interactions between various dissipative mechanisms.

As shown in \ref{app:Green_func}, the three-point correlation
functions can be expressed via three-point retarded Green's functions
as
\bea\label{eq:J_XYZ_final}
\beta^2\!\!\int\! d^4x_1 d^4x_2\Big(\hat{X}(x),\hat{Y}(x_1),\hat{Z}(x_2)\Big)
=-\frac{1}{2}\frac{\partial}{\partial\omega_1}\frac{\partial}{\partial\omega_2}{\rm
  Re}\,G^R_{\hat{X}\hat{Y}\hat{Z}}(\omega_1,\omega_2)\bigg\vert_{\omega_{1,2}=0},
\eea
where 
\bea\label{eq:green_fourier3_final}
G^R_{\hat{X}\hat{Y}\hat{Z}}(\omega_1,\omega_2) 
= -\frac{1}{2}
\int_{-\infty}^{0}\!\! dt_1 e^{-i\omega_1 t_1}\!\!
\int_{-\infty}^{0}\!\! dt_2 e^{-i\omega_2 t_2}\!\!
\int\! d^3x_1\! \int\! d^3x_2\, \nonumber\\ 
\times
\bigg\{
\Big\langle\big[\big[\hat{X}(\bm 0, 0),
\hat{Y}(\bm x_1,t_1)\big],\hat{Z}(\bm x_2,t_2)\big]\Big\rangle_l \nonumber\\
+\Big\langle\big[\big[\hat{X}(\bm 0, 0),
\hat{Z}(\bm x_2,t_2)\big],\hat{Y}(\bm x_1,t_1)\big]\Big\rangle_l \bigg\}
\eea
is the Fourier transform of the three-point retarded correlator taken
in the zero-wavenumber limit. For example, the coefficients
$\lambda_\pi$ which is coupled with the quadratic term
$\sigma_{\rho<\mu}\sigma_{\nu>}^{\rho}$ is given by
\bea\label{eq:lambda_pi1}
\lambda_\pi &=& \frac{12}{35}\beta^2\! \int d^4x_1d^4x_2
\Big(\hat{\pi}_{\gamma}^{\delta}(x),\hat{\pi}_{\delta}^{\lambda}(x_1),\hat{\pi}_{\lambda}^{\gamma}(x_2)\Big)\nonumber\\
&=& -\frac{6}{35}\frac{\partial}{\partial\omega_1}\frac{\partial}{\partial\omega_2}{\rm Re}G^R_{\hat{\pi}_{\gamma}^{\delta} \hat{\pi}_{\delta}^{\lambda} \hat{\pi}_{\lambda}^{\gamma}}(\omega_1,\omega_2)\bigg\vert_{\omega_{1,2}=0}.
\eea

Secondly, additional non-linear terms arise from the non-local
corrections~\eqref{eq:stat_average_21} as a result of the fact that
different dissipative processes become coupled beyond the first-order
as a result of the coupling of two-point correlation functions between
tensors of different rank that are evaluated at distinct space-time
points.  These are the new terms that were omitted in our previous
paper~\cite{Harutyunyan_hydro2022}.  They contain a product or a
contraction of one of the thermodynamic forces $\sigma_{\mu\nu}$,
$\theta$ and $\nabla_\mu\alpha_a$ with the fluid acceleration
$\dot{u}_\mu$, \ie, they will vanish in the case of homogeneous
flow. The terms of this type are multiplied by transport coefficients
which are not independent but are combinations of the relevant
first-order transport coefficients and the relaxation times.

In conclusion, it is important to note that the hydrodynamic equations
can become acausal and unstable due to the second-order terms that are
quadratic in thermodynamic forces
\cite{2012PhRvD..85k4047D,2015JHEP...02..051F}.  One can avoid this
drawback by transforming certain nonlinear terms through the application
of Navier-Stokes equations.  For example,
$\lambda_\pi \sigma_{\rho<\mu} \sigma_{\nu>}^\rho$ can be replaced
with
$\left(\lambda_\pi / 2 \eta\right) \pi_{\rho<\mu} \sigma_{\nu>}^\rho$
\cite{2012PhRvD..85k4047D,2015JHEP...02..051F}. These substitutions
allow our equations to be expressed in a form equivalent to the
complete second-order hydrodynamic equations derived using the method
of moments in Refs~\cite{2011EPJWC..1307005B,Fotakis2022}.

\subsection{Concluding remarks}
\label{sec:hydro_conc}

In this work, utilizing Zubarev's non-equilibrium statistical-operator
formalism, we extend the recent results derived from this
framework~\cite{Harutyunyan_hydro2022} to incorporate additional
second-order terms. These terms arise because Curie's theorem,
  which governs symmetry constraints, can be extended to construct
  invariants from tensors of different ranks evaluated at distinct
  space-time points due to the presence of the acceleration vector in
  the system.

Similar to Ref.~\cite{Harutyunyan_hydro2022}, we focus on a
quantum system characterized by its energy-momentum tensor and the
currents of multiple conserved charges. By employing a second-order
expansion of the statistical operator, we derive complete second-order
equations for the shear stress tensor, bulk viscous pressure, and
flavor diffusion currents.

In particular, we demonstrated that the new additional non-linear
terms in the second-order equations emerge from the memory effects of
the statistical operator and manifest in accelerating relativistic
fluids. These terms capture the non-local mixing between different
dissipative processes in the two-point correlation functions, which
were omitted in the previous analysis~\cite{Harutyunyan_hydro2022}.

Interestingly, although these terms are quadratic in thermodynamic
gradients, they originate from the first-order terms in the Taylor
expansion of the statistical operator. However, deriving them requires
accounting for memory effects and non-locality in the correlation
functions. We also established relations between the new transport
coefficients associated with these terms and the first-order transport
coefficients. Additionally, we established new Kubo-type relations
between the three-point retarded Green's functions and the
second-order transport coefficients which arise from quadratic terms
in the Taylor expansion of the statistical operator.

Looking ahead, it will be important to compute
  second-order transport coefficients for specific systems, both in
  the strong-coupling regime -- where the use of Kubo formulas
  provides a significant advantage -- and in weakly coupled systems,
  where results can be compared against those obtained via
  perturbative methods. The details of such computations will, of
  course, depend on the particular system under consideration. These
  studies are also essential for delineating the domain of
  applicability of second-order relativistic hydrodynamics, which has
  been shown~\cite{Kovtun:2011np} to be limited to frequencies above a
  certain scale -- a limitation that is manifest in the divergences of
  some second-order transport coefficients.

\section*{Acknowledgments}

This research was funded by the Higher Education and Science Committee
(HESC) of the Republic of Armenia through ``Remote Laboratory''
program, Grant No. 24RL-1C010,  Deutsche Forschungsgemeinschaft Grant
No. SE 1836/6-1 and the Polish NCN Grant No. 2023/51/B/ST9/02798.

\appendix

\section{Choice of relevant statistical operator in Zubarev's
  approach}
\label{sec:Relevant_NSO}

The purpose of this Appendix is to clarify the mechanism by which the 
non-hydrodynamical modes are eliminated from the analysis and,
therefore, the potential breakdown of second-order relativistic
hydrodynamics, as argued in Ref.~\cite{Kovtun:2011np}, can be avoided. 

In Zubarev's formalism, one constructs a {\it exact nonequilibrium statistical
operator $\hat{\rho}(t)$} satisfying a modified Liouville equation with
a source term to select retarded solutions
\bea
\frac{d \hat{\rho}(t)}{d t}+i \hat{L} \hat{\rho}(t)
=-\epsilon\left(\hat{\rho}(t)-\hat{\rho}_{\mathrm{rel}}(t)\right),
\eea
where $\hat{L}$ is the Liouville operator, $\hat \rho_{\mathrm{rel}}\equiv \hat{\rho}_{l}$
is the {\it relevant statistical operator}, built from local observables,
such as energy-momentum tensor and currents,
$\epsilon \rightarrow 0^{+}$ ensures proper causal behavior.
It has the formal solution given by
\bea\label{eq:A3}
\hat{\rho}(t)=\epsilon \int_{-\infty}^t
d t^{\prime} e^{\epsilon\left(t^{\prime}-t\right)}
e^{-i\left(t-t^{\prime}\right) \hat{L}} \hat{\rho}_{\mathrm{rel}}\left(t^{\prime}\right),
\eea
 which in the limit $\epsilon \rightarrow 0^{+}$ becomes
\bea
\hat{\rho}(t)=\hat{\rho}_{\rm rel}(t)-\int_{-\infty}^t d t^{\prime}
e^{\epsilon\left(t^{\prime}-t\right)} e^{-i\left(t-t^{\prime}\right)
  \hat{L}}\left(\frac{d}{d t^{\prime}} \hat{\rho}_{\rm rel}\left(t^{\prime}\right)\right).
\eea
Then, the relevant statistical operator
$\hat{\rho}_{\rm rel}(t)$ is given via the maximum entropy
principle, constrained by local expectation values of conserved
quantities such as energy, momentum, charges, etc., with suitable
choice of Lagrange multipliers. This can be formally
written as
\bea
\hat{\rho}_{\rm rel}(t)=\frac{1}{Z_{\rm rel}}
\exp \left[-\int d^3 x \sum_a \lambda_a(\mathbf{x}, t) \hat{P}_a(\mathbf{x})\right],
\eea
where $\hat{P}_a$ stands for local hydrodynamics quantities such as
$T^{\mu \nu}(x)$ and $N^\mu(x)$ and $\lambda_a(\mathbf{x}, t)$ -- for
local Lagrange multipliers, \ie, $\beta(x), \mu(x)$, and $u^\mu(x)$,
and $Z_{\rm rel} $ is the normalization.
Therefore, $\hat{\rho}_{\rm rel}$ resembles formally the
local-equilibrium statistical operator, however the thermodynamical
parameters are space-time-dependent. The coarse-graining inherent to
any hydrodynamical description of matter implies that the space
variations of the thermodynamic parameters obey the inequality
$L\gg \lambda$ with a similar inequality in the time domain.  This
ansatz neglects higher-order (non-hydrodynamic) observables.

To go beyond ideal hydrodynamics, one expands the nonequilibrium
statistical operator $\hat{\rho}(t)$ around the relevant statistical
operator $\hat{\rho}_{\rm rel}(t)$ in spatial and temporal
gradients. This procedure yields first-order (Navier-Stokes) and
second-order corrections. The integral in Eq.~\eqref{eq:A3} then
generates nonlocal contributions expressed through space and time
non-local kernels, through the derivatives of the Lagrange multipliers
$\lambda_a(\mathbf{x}, t)$.

In Zubarev's approach, only hydrodynamic observables, namely,
conserved quantities and their local densities are included in
$\hat{\rho}_{\rm rel}(t)$. This effectively eliminates
non-hydrodynamic modes that originate from higher-order correlation
functions. As a result, the relevant statistical operator
$\hat{\rho}_{\rm rel}(t)$ varies slowly in space and time,
corresponding to the small Knudsen number regime, and the gradient
expansion becomes a systematic framework for deriving relativistic
hydrodynamics. In essence, Zubarev's formalism {\it assumes} that the
macroscopic dynamics is fully determined by the hydrodynamic variables
$\hat{P}_a$, with contributions from higher-order correlators being
negligible.

Consider now, as an illustration, the linear response expression
which arises in the first order as
\bea
\left\langle\hat{P}_a(x)\right\rangle
-\left\langle\hat{P}_a(x)\right\rangle_{\mathrm{rel}}
=\sum_b \int d^4 y \Phi_{a b}(x-y) \nabla \lambda_b(y)+\ldots,
\eea
where $\Phi_{a b}(x-y)$ is a (retarded) correlation function of hydrodynamic operators, \eg,
\bea\label{eq:A6}
\Phi_{a b}(x-y)=\int_0^{\infty}
d \tau\left\langle\left[\hat{P}_a(x),
    \hat{P}_b(y-\tau)\right]\right\rangle_{\mathrm{rel}}.
\eea
Computing $\Phi_{a b}$ from \eqref{eq:A6} presumes that only
hydrodynamic operators matter, \ie, non-hydrodynamic correlators are
neglected.  Thus, in this formulation, it is assumed that the
statistical operator $\rho_{\rm rel}$, constructed from hydrodynamic
fields alone, is sufficient to capture the relevant correlations. This
constitutes a truncation of the hierarchy of dynamical variables,
which is an assumption that plays a role similar to the molecular
chaos hypothesis in kinetic theory. It effectively excludes
non-hydrodynamic fluctuations and long-lived memory effects, thereby
closing the system of equations at the hydrodynamic level.  This
eliminates the need to carry along additional equations for
fluctuating quantities or non-hydrodynamic modes on the scales
relevant to hydrodynamical evolution.

Although our explicit discussion focused on spatial gradients, the
same reasoning applies to temporal gradients. Expanding in time
derivatives implies that correlations decay rapidly, corresponding to
a Markovian approximation. It is important to emphasize that the
NESO formalism does not derive
the form of the relevant statistical operator $\hat{\rho}_{\rm rel}$
from first principles. Instead, its structure is postulated to retain
only the hydrodynamic modes, \ie, the densities of conserved
quantities, which enables a controlled expansion in gradients.

\section{Decomposing the thermodynamic force in different dissipative processes}
\label{app:C_decomp}

For our computations, it is convenient to decompose the operator
$\hat{C}$ given by Eq.~\eqref{eq:C_op} into distinct dissipative
processes using Eqs.~\eqref{eq:T_munu_decomp} and
\eqref{eq:N_a_decomp}.  Recalling the properties
\eqref{eq:orthogonality} and \eqref{eq:prop_proj} we can write
\bea\label{eq:op_C_decompose1}
\hat{C} =\hat{\epsilon} D\beta - \hat{p}
\beta\theta -\sum\limits_a\hat{n}_aD\alpha_a 
+ \hat{q}_{\sigma}(\beta Du^{\sigma}+\nabla^{\sigma}\beta)
-\sum\limits_a\hat{j}^{\sigma}_a \nabla_\sigma\alpha_a
+ \beta\hat{\pi}_{\rho\sigma}\partial^\rho u^\sigma,
\eea
where we used the notations $D = u^{\rho}\partial_{\rho}$,
$\theta = \partial_{\rho}u^{\rho}$,
$\nabla_\sigma= \Delta_{\sigma\rho}\partial^\rho$ introduced in
Sec.~\ref{sec:eq_hydro}.  The initial three terms represent scalar
dissipative processes, the subsequent two terms correspond to vector
dissipative processes, and the final term represents the tensor
dissipative processes.
Next, we have
\bea\label{eq:D_beta}
D\beta &=& \left.\frac{\partial\beta}
{\partial\epsilon}\right|_{n_a}D\epsilon +
\sum\limits_a
\left.\frac{\partial\beta}{\partial n_a}
\right|_{\epsilon, n_b\neq n_a}\!\!\!Dn_a
= -h\theta \left.\frac{\partial\beta}
{\partial\epsilon}\right|_{n_a} -\sum\limits_a
n_a\theta \left.\frac{\partial\beta}{\partial n_a}
\right|_{\epsilon, n_b\neq n_a}\nonumber\\
&&- (\Pi\theta +\partial_\mu q^{\mu}-
q^{\mu}Du_\mu -\pi^{\mu\nu}\sigma_{\mu\nu}) 
\left.\frac{\partial\beta}
{\partial\epsilon}\right|_{n_a} -
\sum\limits_a\partial_\mu j^{\mu}_a 
\left.\frac{\partial\beta}{\partial n_a}
\right|_{\epsilon, n_b\neq n_a},\\
\label{eq:D_alpha}
D\alpha_c &=& \left.\frac{\partial\alpha_c}
{\partial\epsilon}\right|_{n_a} D\epsilon +
\sum\limits_a
\left.\frac{\partial\alpha_c}{\partial n_a}
\right|_{\epsilon, n_b\neq n_a} \!\!\!Dn_a
= -h\theta \left.\frac{\partial\alpha_c}
{\partial\epsilon}\right|_{n_a} -\sum\limits_a
n_a\theta \left.\frac{\partial\alpha_c}{\partial n_a}
\right|_{\epsilon, n_b\neq n_a}\nonumber\\
&& -(\Pi\theta +\partial_\mu q^{\mu}-q^{\mu}Du_\mu 
-\pi^{\mu\nu}\sigma_{\mu\nu})
\left.\frac{\partial\alpha_c}{\partial\epsilon}\right|_{n_a}
 - \sum\limits_a\partial_\mu j^{\mu}_a \left.\frac{\partial \alpha_c}
 {\partial n_a}\right|_{\epsilon, n_b\neq n_a}\!,\qquad\qquad
\eea
where we used Eqs.~\eqref{eq:hydro1} and 
\eqref{eq:hydro2} to eliminate the terms $D\epsilon$, 
$Dn_a$. Now we use the first law of thermodynamics 
and the Gibbs--Duhem relation written as
\bea\label{eq:thermodyn}
ds=\beta d\epsilon -\sum\limits_a\alpha_a dn_a,\qquad
\beta dp  =-hd\beta +\sum\limits_a n_ad\alpha_a.
\eea
We obtain from the first equation the set of Maxwell relations
\bea\label{eq:rel_alpha_beta}
\left.\frac{\partial \beta}{\partial n_a}
\right|_{\epsilon,n_b\neq n_a}=-
\left.\frac{\partial \alpha_a}{\partial\epsilon}
\right|_{n_b},\qquad
\left.\frac{\partial \alpha_c}{\partial n_a}
\right|_{\epsilon, n_b\neq n_a}=
\left.\frac{\partial \alpha_a}{\partial n_c}
\right|_{\epsilon, n_b\neq n_c},
\eea
and from the second equation, we immediately read off
\bea\label{eq:h_n_a}
h=-\beta\left.\frac{\partial p}{\partial\beta}
\right|_{\alpha_a},\qquad
n_a=\beta\left.\frac{\partial p}{\partial\alpha_a}
\right|_{\beta,\alpha_b\neq \alpha_a}.
\eea
Substituting Eqs.~\eqref{eq:rel_alpha_beta} and \eqref{eq:h_n_a} into
the first two terms of Eqs.~\eqref{eq:D_beta} and \eqref{eq:D_alpha}
we obtain
\bea\label{eq:D_beta0}
\beta\theta \left( \left.\frac{\partial p}{\partial\beta}
\right|_{\alpha_a} \left.\frac{\partial\beta}{\partial\epsilon}
\right|_{n_a} +\sum\limits_a
\left.\frac{\partial p}{\partial\alpha_a}
\right|_{\beta,\alpha_b\neq\alpha_a}
\left.\frac{\partial \alpha_a}{\partial\epsilon}\right|_{n_b} 
\right) &=& \beta\theta \gamma,\\
\label{eq:D_alpha0}
-\beta\theta \left( \left.\frac{\partial p}{\partial\beta}
\right|_{\alpha_a} \left.\frac{\partial \beta}{\partial n_c}
\right|_{\epsilon,n_b\neq n_c}+\sum\limits_a
\left.\frac{\partial p}{\partial\alpha_a}
\right|_{\beta,\alpha_b\neq \alpha_a}
\left.\frac{\partial \alpha_a}{\partial n_c}
\right|_{\epsilon, n_b\neq n_c}\right) &=& -\beta\theta\delta_c,
\eea
where
\bea\label{eq:gamma_delta_a}
\gamma \equiv \left.\frac{\partial p}{\partial\epsilon}
\right|_{n_a},\qquad
\delta_a \equiv \left.\frac{\partial p}
{\partial n_a} \right|_{\epsilon, n_b\neq n_a}.
\eea
Then, the first two terms are combined in a single term as above and
we obtain
\bea\label{eq:D_beta1}
D\beta 
&=&\beta\theta \gamma- (\Pi\theta +\partial_\mu q^{\mu}-
q^{\mu}Du_\mu 
-\pi^{\mu\nu}\sigma_{\mu\nu})
\left.\frac{\partial\beta}{\partial\epsilon}\right|_{n_a} -
\sum\limits_a\partial_\mu j^{\mu}_a
\left.\frac{\partial\beta}{\partial n_a}\right|_{\epsilon, n_b\neq n_a},\\
\label{eq:D_alpha1}
D\alpha_c &=& -\beta\theta\delta_c
-(\Pi\theta +\partial_\mu q^{\mu}-q^{\mu}Du_\mu 
-\pi^{\mu\nu}\sigma_{\mu\nu})
\left.\frac{\partial\alpha_c}{\partial\epsilon}\right|_{n_a} -
\sum\limits_a\partial_\mu j^{\mu}_a \left.\frac{\partial
\alpha_c}{\partial n_a}\right|_{\epsilon, n_b\neq n_a}\!.\quad
\eea
Now the first three terms in Eq.~\eqref{eq:op_C_decompose1}
corresponding to scalar dissipation can be combined as follows
\bea\label{eq:scalar_part_2nd}
\hat{\epsilon} D\beta- \hat{p}
\beta\theta -\sum\limits_a \hat{n}_a D\alpha_a  = 
- \beta\theta \hat{p}^*- \hat{\beta}^*(\Pi\theta +\partial_\mu q^{\mu}-
q^{\mu}Du_\mu -\pi^{\mu\nu}\sigma_{\mu\nu})
+\sum\limits_a \hat{\alpha}_a^*\partial_\mu j^{\mu}_a,\quad
\eea
where we exploited the relations~\eqref{eq:rel_alpha_beta}. 

Next we use Eq.~\eqref{eq:hydro3} in the form [the gradient of
pressure is modified according to the second relation in
Eq.~\eqref{eq:thermodyn}]
\bea\label{eq:Du_modify}
h D u_{\sigma}
&=&-hT\nabla_\sigma \beta +
T\sum\limits_a n_a\nabla_\sigma \alpha_a+
\nabla_\sigma\Pi -\Pi D u_\sigma
\nonumber\\
&& - \Delta_{\sigma\mu}D q^{\mu}
-q^{\mu}\partial_\mu u_{\sigma}-q_{\sigma}\theta 
-\Delta_{\sigma\nu}\partial_\mu \pi^{\mu\nu}, 
\eea
to modify the vector term involving $\hat{q}^\sigma$ in
Eq.~\eqref{eq:op_C_decompose1}
\bea\label{eq:vector_part_2nd}
\hat{q}^{\sigma}(\beta Du_{\sigma}+
\nabla_{\sigma}\beta)=\sum\limits_a\frac{n_a}{h}
\hat{q}^{\sigma}\nabla_\sigma \alpha_a -
 \hat{q}^{\sigma}\beta h^{-1}\times \nonumber\\
  (-\nabla_\sigma \Pi
+\Pi Du_\sigma + D q_{\sigma}
 + q^{\mu}\partial_\mu u_{\sigma} +q_{\sigma}\theta
  +\partial_\mu \pi^{\mu}_{\sigma}).
\eea
Combining Eqs.~\eqref{eq:op_C_decompose1}, \eqref{eq:scalar_part_2nd},
and \eqref{eq:vector_part_2nd} and replacing
$\partial^{\rho}u^{\sigma} \to
\sigma^{\rho\sigma}=\Delta^{\rho\sigma}_{\mu\nu}\partial^{\mu}u^{\nu}$
in the last term in Eq.~\eqref{eq:op_C_decompose1} according to the
symmetry properties of the shear stress tensor
$\hat{\pi}_{\rho\sigma}$ we obtain the final form of the operator
$\hat{C}$ given in the main text by
Eqs.~\eqref{eq:C_decompose_2nd}--\eqref{eq:op_C2}.

We turn to the computation of operator $\partial_\tau \hat{C}$.
 Using the relation $\partial_\tau \Delta_{\gamma\delta} =
-u_\gamma\partial_\tau u_\delta -u_\delta\partial_\tau u_\gamma$, 
 from Eq.~\eqref{eq:projector_delta4} we obtain
\bea\label{eq:delta_deriv1}
\partial_\tau\Delta_{\gamma\delta\rho\sigma}
&=&-\frac{1}{2}\Big[\Delta_{\gamma\rho}(u_\sigma\partial_\tau 
u_\delta+u_\delta\partial_\tau u_\sigma)+\Delta_{\delta\sigma}
(u_\gamma\partial_\tau u_\rho+u_\rho\partial_\tau u_\gamma)+
(\rho\leftrightarrow\sigma)\Big]\nonumber\\&& +\frac{1}{3}
\Big[\Delta_{\gamma\delta}(u_\rho\partial_\tau u_\sigma+
u_\sigma\partial_\tau u_\rho)+\Delta_{\rho\sigma}
(u_\gamma\partial_\tau u_\delta+u_\delta\partial_\tau u_\gamma)\Big],
\eea
which we will utilize below.

Using the decompositions~\eqref{eq:T_munu_decomp} and 
\eqref{eq:N_a_decomp} we find from Eq.~\eqref{eq:C_op}
\bea\label{eq:C_deriv}
\partial_\tau \hat{C}=\hat{T}_{\rho\sigma}\,
\partial_{\tau}\partial^{\rho}(\beta u^\sigma) -
\sum\limits_a\hat{N}^\rho_a\, \partial_{\tau}\partial_\rho\alpha_a
=(\hat{\epsilon} u_{\rho}u_{\sigma} - \hat{p}\Delta_{\rho\sigma} + 
\hat{q}_{\rho}u_{\sigma}+ \hat{q}_{\sigma}u_{\rho} + 
\hat{\pi}_{\rho\sigma})\nonumber\\
\times \big[\beta \partial_{\tau}\partial^{\rho}u^\sigma +
u^\sigma\partial_{\tau}\partial^{\rho}\beta +
(\partial_{\tau}\beta)(\partial^{\rho} u^\sigma) 
+(\partial_{\tau}u^\sigma)(\partial^{\rho} \beta)\big]
- \sum\limits_a (\hat{n}_a u^\rho +\hat{j}^{\rho}_a) 
\partial_{\tau}\partial_\rho\alpha_a\nonumber\\
=\hat{\epsilon} X^{(\epsilon)}_{\tau}-\hat{p} X^{(p)}_{\tau}+
\hat{q}_{\rho} X^{\rho (q)}_{\tau}+\hat{\pi}_{\rho\sigma} 
X^{\rho\sigma(\pi)}_{\tau}- \sum\limits_a 
(\hat{n}_a X^{(n_a)}_{\tau} +\hat{j}_{\rho a} X^{\rho (j_a)}_{\tau}),
\eea
where the corresponding thermodynamic forces are given by 
\bea\label{eq:X_ep}
X_{\tau}^{(\epsilon)} &=& \beta u^{\rho}u^\mu\partial_{\tau}
(\partial_\mu u_\rho) +u^\mu(\partial_{\tau}\partial_\mu\beta)
=\gamma\partial_{\tau}(\beta\theta) 
+\beta\theta(\partial_{\tau}\gamma)
-(\partial_{\tau}u_{\rho})\sum_a \frac{n_a}{h}
\left(\nabla^\rho\alpha_a\right),\nonumber\\\\
\label{eq:X_p}
X_{\tau}^{(p)} &=& 
\beta \partial_{\tau}\theta
-\beta u_{\rho}
u_{\sigma}(\partial_{\tau}\partial^{\rho}u^\sigma)
+\theta\partial_{\tau}\beta
+(\partial_{\tau}u^\rho)(\partial_{\rho} \beta)
= \partial_{\tau}(\beta\theta) 
+\partial_\tau u_\rho \sum_a \frac{n_a}{h}
\left(\nabla^\rho\alpha_a\right),\nonumber\\\\
\label{eq:X_pi}
X_{\tau}^{\rho\sigma (\pi)} &= & \partial_{\tau}(\beta \partial^{\rho}u^\sigma) 
+(\partial_{\tau}u^\sigma)(\partial^{\rho} \beta)
=\partial_{\tau} (\beta \sigma^{\rho\sigma})
+(\partial_{\tau}u^\rho) \sum_a \frac{n_a}{h}\nabla^\sigma\alpha_a,
\eea
\bea\label{eq:X_q}
X_{\tau}^{\rho(q)} &=&
\beta
u_{\sigma}(\partial_{\tau}\partial^{\rho}u^\sigma)
+\partial_{\tau}\partial^{\rho}\beta +
\beta D\partial_{\tau}u^\rho +
(\partial_{\tau}\beta)(D u^\rho) 
+(\partial_{\tau}u^\rho)(D \beta)\nonumber\\
&= & -2\beta \sigma^{\rho\sigma}(\partial_\tau u_\sigma) 
-2\left(\frac{1}{3}-\gamma\right)\beta \theta(\partial_\tau u^\rho) 
+\sum_a \frac{n_a}{h}\partial_{\tau}\left(\nabla^\rho\alpha_a\right)
+\sum_a \partial_{\tau}(n_ah^{-1})\nabla^\rho\alpha_a,\nonumber\\\\
\label{eq:X_na}
X_{\tau}^{(n_a)} &=& 
u^\mu(\partial_{\tau}\partial_\mu\alpha_a)
=-\delta_a\partial_{\tau}(\beta\theta)
-\beta\theta(\partial_{\tau}\delta_a)
-(\partial_{\tau}u^\rho)(\nabla_\rho\alpha_a),\\
\label{eq:X_ja}
X_{\tau}^{\rho (j_a)} &=& \partial_{\tau}(\partial^\rho\alpha_a)
= \partial_{\tau}(\nabla^\rho\alpha_a)-
(\beta\theta\delta_a) \partial_{\tau}u^\rho,
\eea
where we used Eqs.~\eqref{eq:D_beta1}, \eqref{eq:D_alpha1} 
and \eqref{eq:Du_modify} and dropped the second-order corrections.
 In Eqs.~\eqref{eq:X_pi},
\eqref{eq:X_q} and \eqref{eq:X_ja} we used the orthogonality
properties~\eqref{eq:orthogonality}. In addition, we 
used Eq.~\eqref{eq:delta_deriv1} in 
Eq.~\eqref{eq:X_pi} and dropped the terms 
$\propto u^\rho, u^\sigma, \Delta^{\rho\sigma}$ which are
orthogonal to $\hat{\pi}_{\rho\sigma}$.

Substituting these expressions back to Eq.~\eqref{eq:C_deriv} 
we obtain
\bea\label{eq:C_deriv1}
\partial_\tau \hat{C}
=-\hat{p}^*\partial_{\tau}(\beta\theta)
+\beta\theta \left[\hat{\epsilon} 
(\partial_{\tau}\gamma) + \sum\limits_a \hat{n}_a 
(\partial_{\tau}\delta_a)\right] - (\partial_{\tau}u_{\rho})
\sum_a \left(\nabla^\rho\alpha_a\right) 
\left[\frac{n_a}{h}(\hat{\epsilon}+\hat{p}) -\hat{n}_a\right]
\nonumber\\
+ \hat{q}_{\rho}\left[
-2\beta \sigma^{\rho\sigma}(\partial_\tau u_\sigma) 
-2\left(\frac{1}{3}-\gamma\right)\beta \theta(\partial_\tau u^\rho) 
+\sum_a \partial_{\tau}(n_ah^{-1})\nabla^\rho\alpha_a
+ \beta\theta (\partial_{\tau}u^\rho) h^{-1}
\sum\limits_a \delta_a n_a \right]\nonumber\\
+ \sum\limits_a \hat{\mathscr{J}}^{\rho}_a
\Big[\beta\theta (\partial_{\tau}u_\rho) \delta_a 
- \partial_{\tau}(\nabla_\rho\alpha_a) \Big]
+ \hat{\pi}_{\rho\sigma}\left[
\partial_{\tau} (\beta \sigma^{\rho\sigma})
+(\partial_{\tau}u^\rho) \sum_a \frac{n_a}{h}
\nabla^\sigma\alpha_a\right].\qquad
\eea
Note, that in the operator $\partial_\tau \hat{C}$
the thermodynamic forces are taken at the point $x$,
whereas the operators are taken at the point $x_1$.

\section{Correlation functions and Kubo formulas}
\label{app:Green_func}

In this Appendix, we provide the details of deriving the Kubo
relations for the first-order and second-order transport
coefficients. \ref{app:2-point_cor} recapitulates, for the sake of
completeness, Appendix~C of Ref.~\cite{Harutyunyan_hydro2022} which
closely follows similar derivations in
Refs.~\cite{1984AnPhy.154..229H,2011AnPhy.326.3075H}, and
\ref{app:3-point_cor} provides a new derivation of the Kubo-type
formulas for the second-order transport coefficients.  In evaluating
transport coefficients, variations in thermodynamic parameters can be
disregarded. The local equilibrium distribution can be approximated by
a global equilibrium distribution characterized by an average
temperature $T=\beta^{-1}$ and mean chemical potentials $\mu_a$.

\subsection{2-point correlation functions}
\label{app:2-point_cor}

Consider a generic two-point correlator given by
Eq.~\eqref{eq:2_point_corr}.  In full thermal equilibrium the system
is described with the grand canonical distribution with
$\hat{A} =\beta\hat{K}$ in Eq.~\eqref{eq:A_op} with
$\hat{K}=\hat{H}-\sum\limits_a\mu_a\hat{\cal N}_a$ (in the fluid rest
frame) which gives
\bea\label{eq:2_point_corr1}
\Big(\hat{X}(\bm x, t),\hat{Y}(\bm x_1, t_1)\Big)=
\int_0^1\!\! d\lambda \Big\langle\hat{X}(\bm x, t)
\left[e^{-\beta\lambda \hat{K}}\hat{Y}(\bm x_1,t_1)
e^{\beta\lambda \hat{K}} - \big\langle \hat{Y}(\bm
x_1,t_1)\big\rangle_l\right]\Big\rangle_l.
\eea
In the Heisenberg picture, the time evolution of any operator follows the expression
\bea\label{eq:op_heisenberg}
\hat{Y}(\bm x,t) = e^{i\hat{K}t} \hat{Y}(\bm x,0) e^{-i\hat{K}t}.
\eea
Consequently, by shifting the time variable by an infinitesimal amount, we obtain
\bea
\hat{Y}(\bm x,t+\delta t) = e^{i\hat{K} (t+\delta t)} \hat{Y}(\bm x,0)
e^{-i\hat{K}(t+\delta t)} = e^{i\hat{K} \delta t} \hat{Y}(\bm x,t) e^{-i\hat{K} \delta t}.
\eea
By performing an analytic continuation, specifically setting $\delta t
\rightarrow i \tau$,  we arrive at
\bea\label{eq:op_heisenberg1}
\hat{Y}(\bm x,t+i\tau) = e^{-\hat{K}\tau} \hat{Y}(\bm x,t) e^{\hat{K}\tau}.
\eea
From this, it follows that
\bea\label{eq:Y_l}
\big\langle \hat{Y}(\bm x,t+i\tau)\big\rangle_l &=& \big\langle \hat{Y}(\bm x,t)\big\rangle_l,\\
\label{eq:KMS}
\big\langle \hat{X}(\bm x, t) \hat{Y} (\bm x_1,t'+i\beta)\big\rangle_l &=& \big\langle\hat{Y}(\bm x_1,t') \hat{X}(\bm x, t) \big\rangle_l.
\eea
The identity in Eq.~\eqref{eq:KMS} corresponds to the well-known
Kubo–Martin–Schwinger (KMS) condition.

Performing a variable change $\tau= \lambda\beta$ in
Eq.~\eqref{eq:2_point_corr1} and employing
Eqs.~\eqref{eq:op_heisenberg1} and \eqref{eq:Y_l} we obtain
\bea\label{eq:2_point_corr2}
\Big(\hat{X}(\bm x, t),\hat{Y}(\bm x_1, t_1)\Big)
= \frac{1}{\beta} \int_0^\beta\!\! d\tau
\Big\langle\hat{X}(\bm x, t)\left[\hat{Y}(\bm x_1,t_1+i\tau)
- \big\langle \hat{Y}(\bm x_1,t_1+i\tau)
\big\rangle_l\right]\Big\rangle_l.
\eea
Assuming that the correlations vanish in the limit
$t_1\to -\infty$~\cite{1984AnPhy.154..229H,2011AnPhy.326.3075H}, \ie,
\bea\label{eq:corr_vanish2}
\lim_{t_1\to -\infty}\left(\big\langle\hat{X}(\bm x, t)\hat{Y}(\bm x_1,t_1+i\tau)\big\rangle_l -
\big\langle\hat{X}(\bm x, t) \big\rangle_l 
\big\langle\hat{Y}(\bm x_1,t_1+i\tau)\big\rangle_l\right)=0,
\eea
we can modify the integrand in Eq.~\eqref{eq:2_point_corr2} as follows 
\bea\label{eq:itgr_mod2}
&&\hspace{2cm}\big\langle\hat{X}(\bm x, t)
\hat{Y}(\bm x_1,t_1+i\tau)\big\rangle_l-
\big\langle\hat{X}(\bm x, t)\big\rangle_l
\big\langle \hat{Y}(\bm x_1,t_1+i\tau)\big\rangle_l \nonumber\\
&&=\Big\langle\hat{X}(\bm x, t) \int_{-\infty}^{t_1}\! dt' 
\frac{d}{dt'}\hat{Y}(\bm  x_1,t'+i\tau)\Big\rangle_l -
\big\langle\hat{X}(\bm x, t)\big\rangle_l
\int_{-\infty}^{t_1}\! dt' \frac{d}{dt'}
\big\langle \hat{Y}(\bm  x_1,t'+i\tau)\big\rangle_l \nonumber\\
&&= -i\int_{-\infty}^{t_1}\! dt' 
\big\langle\hat{X}(\bm x, t)\frac{d}{d\tau}
\hat{Y}(\bm  x_1,t'+i\tau)\big\rangle_l +
i\int_{-\infty}^{t_1}\! dt' 
\big\langle\hat{X}(\bm x, t)\big\rangle_l\frac{d}{d\tau'}
\big\langle \hat{Y}(\bm  x_1,t'+i\tau)\big\rangle_l.\nonumber
\eea
Plugging this result back into Eq.~\eqref{eq:2_point_corr2}  and
applying the relations given by  Eqs.~\eqref{eq:Y_l}  and \eqref{eq:KMS} 
we obtain the final expression (note that the integration over $\tau$
effectively cancels the differentiation, while the second KMS relation
swaps the operators, leading to the commutator)
\bea\label{eq:2_point_corr3}
\Big(\hat{X}(\bm x, t),\hat{Y}(\bm x_1, t_1)\Big)
=\frac{i}{\beta} \int_{-\infty}^{t_1}\! dt' 
\big\langle\big[\hat{X}(\bm x, t),
\hat{Y}(\bm  x_1,t')\big]\big\rangle_l.
\eea
Here, the square brackets represent the commutator.
Considering the time ordering $t^{\prime} \leq t_1 \leq t$, we can express
Eq.~\eqref{eq:2_point_corr3} as follows:
\bea\label{eq:2_point_corr5}
\Big(\hat{X}(\bm x, t),\hat{Y}(\bm x_1, t_1)\Big)
= -\frac{1}{\beta} \int_{-\infty}^{t_1}\! dt'\, 
G^R_{\hat{X}\hat{Y}}(\bm x-\bm x_1, t-t'),
\eea
where 
\bea\label{eq:retarded_corr}
G^R_{\hat{X}\hat{Y}}(\bm x-\bm x', t-t')=
-i\theta(t-t')\big\langle\big[\hat{X}(\bm x, t),
\hat{Y}(\bm  x',t')\big]\big\rangle_l
\eea
is the retarded two-point Green's function for a uniform medium.

Now consider a generic transport coefficient given by the integral
\bea\label{eq:I_XY}
I[{\hat{X},\hat{Y}}](\omega) = \beta\!\int\! d^3x_1\!
\int_{-\infty}^t\!\! dt_1 e^{i\omega(t-t_1)}
e^{\varepsilon(t_1-t)}\Big(\hat{X}(\bm x, t),\hat{Y}
(\bm x_1, t_1)\Big),
\eea
For convenience, we also introduce a nonzero frequency $\omega>0$; the
limit $\omega \rightarrow 0$ will be taken at the final stage of the
calculations.  According to Eq.~\eqref{eq:2_point_corr5} we can write
Eq.~\eqref{eq:I_XY} as
\bea\label{eq:I_XY1}
I[{\hat{X},\hat{Y}}](\omega)
=-\int_{-\infty}^0\! dt' e^{(\varepsilon-i\omega) t'}\!\!
\int_{-\infty}^{t'}\! dt\! \int\! d^3x\,
G^R_{\hat{X}\hat{Y}}(-\bm x, -t).
\eea
Considering the Fourier transformation
\bea\label{eq:green_fourier1}
G^R_{\hat{X}\hat{Y}}(\bm x, t) =
\int\! \frac{d^3k}{(2\pi)^3}\int_{-\infty}^{\infty} \frac{d\omega'}{2\pi}e^{-i(\omega' t-\bm k\cdot \bm x)}
G^R_{\hat{X}\hat{Y}}(\bm k,\omega'),\nonumber
\eea
we obtain
\bea
\int\! d^3x\, G^R_{\hat{X}\hat{Y}}(-\bm x, -t)=
\int_{-\infty}^{\infty} \frac{d\omega'}{2\pi}
e^{i\omega' t}G^R_{\hat{X}\hat{Y}}(\omega'),\nonumber
\eea
where
$G^R_{\hat{X}\hat{Y}}(\omega')\equiv\lim_{\bm k\to 0}
G^R_{\hat{X}\hat{Y}}(\bm k,\omega')$.  In Eq.~\eqref{eq:I_XY1} we now
encounter the integral $\int_{-\infty}^{t'} dt\, e^{i\omega' t}$,
which we compute by a shift $\omega' \rightarrow \omega' - i \delta$,
$\delta >0$, taking the limit $\delta \rightarrow 0^+$ at the end
(this modification is called regularization and helps us handle the
potential divergence at $t \rightarrow-\infty$)
\bea
\int_{-\infty}^{t'} dt\, e^{i\omega' t}
= \lim_{\delta \to 0^+} \int_{-\infty}^{t'} dt\, e^{(i\omega' + \delta)t}
= \lim_{\delta \to 0^+} \frac{e^{(i\omega' + \delta) t'}}{i \omega' + \delta}.
\eea
The factor $e^{(i \omega'+\delta) t^{\prime}}$ ensures that the
integral converges as $t^{\prime} \rightarrow-\infty$, as long as
$\delta>0$. Therefore,
\bea\label{eq:exp_integral}
\int_{-\infty}^0\! dt' e^{(\varepsilon-i\omega) t'}\! \int_{-\infty}^{t'} dt\, e^{i\omega' t}
= \lim_{\delta \to 0^+} \int_{-\infty}^0\! dt' 
e^{(\varepsilon-i\omega) t'}
\frac{e^{(i\omega'+ \delta) t'}}{i\omega'+ \delta} \nonumber\\
=-\lim_{\delta \to 0^+}
\frac{1}{[\omega' -\omega-i(\varepsilon+ \delta)](\omega'-i \delta)}\nonumber\\
=-\lim_{\delta \to 0^+}\frac{1}{\omega+i\varepsilon}
\left(\frac{1}{\omega' -\omega-i(\varepsilon+ \delta)}
  -\frac{1}{\omega'-i \delta}\right).
\eea
Then we have from Eq.~\eqref{eq:I_XY1} 
\bea\label{eq:I_XY2}
I[{\hat{X},\hat{Y}}] (\omega)
&=&-\lim_{\delta \to 0^+} \int_{-\infty}^{\infty}\frac{d\omega'}{2\pi}\,
G^R_{\hat{X}\hat{Y}}(\omega')
\!\int_{-\infty}^0\! dt' 
e^{(\varepsilon-i\omega) t'}
\frac{e^{(i\omega' + \delta) t'}}{i\omega' + \delta}\nonumber\\
&=&\lim_{\delta \to 0^+}\frac{i}{\omega+i\varepsilon}
\oint\frac{d\omega'}{2\pi i}
\left(\frac{1}{\omega' -\omega-i(\varepsilon+ \delta)}
  -\frac{1}{\omega'-i \delta}\right)G^R_{\hat{X}\hat{Y}}(\omega').\nonumber
\eea
In this case, the contour integral is closed in the upper half-plane,
where the retarded Green's function is analytic. The contribution from
the semicircle at infinity vanishes as long as the retarded Green's
function decays sufficiently quickly, specifically no slower than
$\omega^{-1}$, which we assume is true in this context.

By applying Cauchy's integral formula and taking the
limits $\delta \rightarrow 0^{+}$and $\varepsilon \rightarrow 0^{+}$,
we obtain the following result:
\bea\label{eq:I_XY3}
I[{\hat{X},\hat{Y}}] (\omega)
= \frac{i}{\omega}\left[G^R_{\hat{X}\hat{Y}}
(\omega)-G^R_{\hat{X}\hat{Y}}(0)\right].
\eea
Going to the zero-frequency limit $\omega\to 0$ we obtain the final
formula
\bea\label{eq:I_XY4}
I[{\hat{X},\hat{Y}}](0)=i\frac{d}{d\omega}
G^R_{\hat{X}\hat{Y}}(\omega)\bigg\vert_{\omega=0},
\eea
with
\bea\label{eq:green_fourier2}
G^R_{\hat{X}\hat{Y}}(\omega) 
= -i\int_{0}^{\infty}\!\! dt e^{i\omega t}\!\!
\int\! d^3x\, \big\langle\big[\hat{X}(\bm x, t),
\hat{Y}(\bm 0,0)\big]\big\rangle_l.
\eea
From Eqs.~\eqref{eq:green_fourier2} and \eqref{eq:I_XY3} we find that
\bea\label{eq:parity_green}
\big\{G^R_{\hat{X}\hat{Y}}(\omega)\big\}^* =
G^R_{\hat{X}\hat{Y}}(-\omega),\qquad
\big\{I[\hat{X},\hat{Y}](\omega)\big\}^* =
I[\hat{X},\hat{Y}](-\omega).
\eea
Indeed, since $\hat{X}(\bm x, t)$ and 
$\hat{Y}(\bm x, t)$
are hermitian operators, we have the property
\bea
\big\langle\big[\hat{X}(\bm x, t),
\hat{Y}(\bm  x',t')\big]\big\rangle_l^*=
-\big\langle\big[\hat{X}(\bm x, t),
\hat{Y}(\bm  x',t')\big]\big\rangle_l,
\eea
therefore the retarded Green's function given by
Eq.~\eqref{eq:retarded_corr} is real, which is used to obtain the
first relation in Eq.~\eqref{eq:parity_green}. From
Eq.~\eqref{eq:parity_green} we have also
\bea\label{eq:parity_green1}
{\rm Re}G^R_{\hat{X}\hat{Y}}(-\omega)=
{\rm Re}G^R_{\hat{X}\hat{Y}}(\omega),\qquad
{\rm Im}G^R_{\hat{X}\hat{Y}}(-\omega)=
-{\rm Im}G^R_{\hat{X}\hat{Y}}(\omega),
\eea
therefore from Eqs.~\eqref{eq:I_XY} and \eqref{eq:I_XY4} we obtain in
the zero-frequency limit
\bea\label{eq:I_XY5}
I[{\hat{X},\hat{Y}}](0)= \beta\!\int\! d^4x_1
\Big(\hat{X}(x),\hat{Y}(x_1)\Big) =-\frac{d}{d\omega}
{\rm Im}G^R_{\hat{X}\hat{Y}}(\omega)\bigg\vert_{\omega=0},
\eea
where we used the short-hand notation 
defined in Eq.~\eqref{eq:int_short}.

Now let us show that the Green's function \eqref{eq:green_fourier2} is
symmetric in its arguments if the operators $\hat{X}$ and $\hat{Y}$
have the same parity under time reversal. We have
\bea\label{eq:Green_trans}
G^R_{\hat{Y}\hat{X}}(\omega) 
&=& i\!\int_{0}^{\infty}\!\! dt e^{i\omega t}\!\! 
\int\! d^3x\,\big\langle\big[\hat{X}(\bm 0,0),
\hat{Y}(\bm x,t)\big]\big\rangle_l\nonumber\\
&=& i\!\int_{0}^{\infty}\!\! dt e^{i\omega t}\!\!
\int\! d^3x\,\big\langle\big[\hat{X}(-\bm x,-t),
\hat{Y}(\bm 0,0)\big]\big\rangle_l \nonumber\\
&=& i\!\int_{0}^{\infty}\!\! dt e^{i\omega t}\!\!
\int d^3x\,\big\langle\big[\hat{X}(\bm x,-t),
\hat{Y}(\bm 0,0)\big]\big\rangle_l,
\eea
where we used the uniformity of the medium.

For Hermitian operators, their transformation under time reversal is given by
\bea\label{eq:time_reversal}
\hat{X}_T(\bm x, t) = \eta_X \hat{X}(\bm x, -t),
\qquad \hat{Y}_T(\bm x, t) = \eta_Y \hat{Y}(\bm x, -t),\nonumber
\eea
where $\eta_{X, Y}= \pm 1$ depending on whether the operators are even
or odd under time reversal.  Applying this to
Eq.~\eqref{eq:Green_trans}, we obtain
\bea
G^R_{\hat{Y}\hat{X}}(\omega) &=&
i\eta_X\eta_Y \int_{0}^{\infty} dt e^{i\omega t}
\int d^3x\big\langle\big[\hat{X}_T(\bm x,t),
\hat{Y}_T(\bm 0,0)\big]\big\rangle_l\nonumber\\
&=& i\eta_X\eta_Y \int_{0}^{\infty} dt e^{i\omega t}
\int d^3x \big\langle\big[\hat{X}(\bm x,t),
\hat{Y}(\bm 0,0)\big]\big\rangle_{l,T}.\nonumber
\eea

Taking into account that the expectation value of a commutator of
Hermitian operators is purely imaginary and that time reversal is an
antiunitary transformation (which conjugates complex numbers), we
arrive at
\bea\label{eq:green_symmetry}
G^R_{\hat{Y}\hat{X}}(\omega) = -i\eta_X\eta_Y \int_{0}^{\infty}
dt e^{i\omega t} \int d^3x \big\langle\big[\hat{X}(\bm x,t),
\hat{Y}(\bm 0,0)\big]\big\rangle_l =
\eta_X\eta_Y G^R_{\hat{X}\hat{Y}}(\omega).
\eea
Thus, when $\eta_X=\eta_Y$, we obtain the relation
$
G_{\hat{Y} \hat{X}}^R(\omega)=G_{\hat{X} \hat{Y}}^R(\omega)
$
which corresponds to Onsager's symmetry principle for transport
coefficients. Finally, using Eq.~\eqref{eq:I_XY5} along with the
transport coefficient definitions given in Eqs~\eqref{eq:shear_def},
\eqref{eq:bulk_def}, \eqref{eq:chi_ab_def}, and \eqref{eq:kappa_def}
we derive the expressions \eqref{eq:shear_bulk_mod} and
\eqref{eq:kappa_mod} presented in the main text.

In the derivation of the second-order equations of motion for the
dissipative currents we encounter integrals of the type
\bea\label{eq:I_XY_tau}
I^\tau[{\hat{X},\hat{Y}}](\omega)
= \beta\! \int\! d^4x_1 e^{i\omega(t-t_1)}
\Big(\hat{X}(x),\hat{Y}(x_1)\Big)(x_1-x)^\tau,
\eea
where, we again employ the shorthand notation
Eq.~\eqref{eq:int_short}. The correlator
$\left(\hat{X}(x), \hat{Y}\left(x_1\right)\right)$ evaluated in the
local rest frame depends on the spatial coordinates only through the
difference $\left|\boldsymbol{x}-\boldsymbol{x}_1\right|$, meaning it
is an even function of $\boldsymbol{x}-\boldsymbol{x}_1$.

Then Eq.~\eqref{eq:I_XY_tau} implies that the spatial components
of the vector $I^\tau$ vanish in that frame, and for the temporal
component we have
\bea\label{eq:vector_I0_rest}
I^0[{\hat{X},\hat{Y}}](\omega) &=& \beta\! 
\int\! d^4x_1 e^{i\omega(t-t_1)} \Big(\hat{X}(x),
\hat{Y}(x_1)\Big)(t_1-t)\nonumber\\
&=& i\beta\frac{d}{d\omega}\!\int\! d^4x_1
e^{i\omega(t-t_1)}\Big(\hat{X}(x),\hat{Y}(x_1)\Big)
= i\frac{d}{d\omega}I[{\hat{X},\hat{Y}}](\omega),
\eea
where we used Eq.~\eqref{eq:I_XY}.  From Eqs.~\eqref{eq:I_XY3} and
\eqref{eq:vector_I0_rest} we obtain in the limit $\omega\to 0$
\bea \label{eq:vec_I}
I^0[{\hat{X},\hat{Y}}](0) =K[{\hat{X},\hat{Y}}],
\eea
where we defined
\bea\label{eq:K_XY}
K[\hat{X},\hat{Y}] 
 \equiv -\frac{1}{2}\frac{d^2}{d\omega^2} G^R_{\hat{X}
\hat{Y}}(\omega)\bigg\vert_{\omega=0}
 =-\frac{1}{2}\frac{d^2}{d\omega^2} {\rm Re}G^R_{\hat{X}
\hat{Y}}(\omega)\bigg\vert_{\omega=0}.
\eea
It is important to emphasize that in Eqs.~\eqref{eq:I_XY4} and
\eqref{eq:K_XY}, the Green's function must be computed in the fluid
rest frame.

The relation \eqref{eq:vec_I} can also be cast into a covariant form
\bea\label{eq:vec_gen}
\beta\!\int\! d^4x_1 \Big(\hat{X}(x),\hat{Y}
(x_1)\Big)(x_1-x)^\tau =K[{\hat{X},\hat{Y}}]u^\tau.
\eea

\subsection{3-point correlation functions}
\label{app:3-point_cor}

Consider now a generic three-point correlator given by
Eq.~\eqref{eq:3_point_corr}.  Recalling again the definition of
$\hat{X}_\lambda$ after Eq.~\eqref{eq:rho_2_final}, performing
variable change $\beta\lambda_1 =\tau_1$, $\beta\lambda_2 =\tau_2$ and
using Eq.~\eqref{eq:op_heisenberg1} we obtain
\bea\label{eq:3_point_corr1}
\Big(\hat{X}(x),\hat{Y}(x_1),\hat{Z}(x_2)\Big) &=&
\frac{1}{2\beta^2} \int_0^\beta \!  d\tau_1\! \int_0^\beta\! d\tau_2 \, \Big\langle
\hat{X}(\bm x,t)\Big[ \tilde{T}\, \hat{Y}(\bm x_1,t_1+i\tau_1)
\hat{Z}(\bm x_2,t_2+i\tau_2)  \nonumber\\
&& \hspace{-1.7cm} 
 -\big\langle \hat{Y}(\bm x_1,t_1+i\tau_1)\big\rangle_l  
 \hat{Z}(\bm x_2,t_2+i\tau_2)- \hat{Y}(\bm x_1,t_1+i\tau_1) 
\big\langle \hat{Z}(\bm x_2,t_2+i\tau_2) \big\rangle_l \nonumber\\
&& \hspace{-3.3cm}
-\big\langle\tilde{T}\,\hat{Y}(\bm x_1,t_1+i\tau_1)\hat{Z}(\bm x_2,t_2+i\tau_2) \big\rangle_l+
2\big\langle \hat{Y}(\bm x_1,t_1+i\tau_1)\big\rangle_l
\big\langle \hat{Z}(\bm x_2,t_2+i\tau_2) \big\rangle_l\Big]\Big\rangle_l\nonumber\\
&=&
\frac{1}{2\beta^2} \int_0^\beta \!  d\tau_1\! \int_0^\beta\! d\tau_2 \, I\Big((\bm x,t); (\bm x_1,t_1+i\tau_1); (\bm x_2,t_2+i\tau_2)\Big),
\eea
 where on using Eq.~\eqref{eq:Y_l} the integrand can be written in
the form
\bea\label{eq:integrand_3_point}
I\Big((\bm x,t); (\bm x_1,t_1+i\tau_1); (\bm x_2,t_2+i\tau_2)\Big) &=&
 \Big\langle
\hat{X}(\bm x,t)\, \tilde{T}\, \hat{Y}(\bm x_1,t_1+i\tau_1)
\hat{Z}(\bm x_2,t_2+i\tau_2)\Big\rangle_l  \nonumber\\
&&  
 -\big\langle \hat{Y}(\bm x_1,t_1)\big\rangle_l  
 \Big\langle \hat{X}(\bm x,t) \hat{Z}(\bm x_2,t_2+i\tau_2)\Big\rangle_l\nonumber\\
&&
 -\Big\langle \hat{X}(\bm x,t) \hat{Y}(\bm x_1,t_1+i\tau_1) \Big\rangle_l\big\langle \hat{Z}(\bm x_2,t_2) \big\rangle_l \nonumber\\
&& 
- \big\langle \hat{X}(\bm x,t)\big\rangle_l \Big\langle\tilde{T}\,\hat{Y}(\bm x_1,t_1+i\tau_1)\hat{Z}(\bm x_2,t_2+i\tau_2) \Big\rangle_l
\nonumber\\
&&
+2\big\langle \hat{X}(\bm x,t)\big\rangle_l
\big\langle \hat{Y}(\bm x_1,t_1)\big\rangle_l
\big\langle \hat{Z}(\bm x_2,t_2) \big\rangle_l.
\eea

As in the case of two-point correlators, we assume that the
correlations vanish in the limit $t_1,t_2\to -\infty$, \ie,
\bea\label{eq:corr_vanish3}
\lim_{t_1,t_2\to -\infty}
I\Big((\bm x,t); (\bm x_1,t_1+i\tau_1); (\bm x_2,t_2+i\tau_2)\Big)=0.
\eea
Then we can modify the integrand in Eq.~\eqref{eq:3_point_corr1} as
follows
\bea\label{eq:itgr_mod2}
&& I\Big((\bm x,t); (\bm x_1,t_1+i\tau_1); (\bm x_2,t_2+i\tau_2)\Big)\nonumber\\
&& 
=\int_{-\infty}^{t_2}\! dt'' \frac{d}{dt''}
I\Big((\bm x,t); (\bm x_1,t_1+i\tau_1); (\bm x_2,t''+i\tau_2)\Big)\nonumber\\
&& 
=-i\frac{d}{d\tau_2} \int_{-\infty}^{t_2}\! dt'' I\Big((\bm x,t); (\bm x_1,t_1+i\tau_1); (\bm x_2,t''+i\tau_2)\Big),
\eea 
which gives
\bea\label{eq:3_point_corr2}
\int_0^\beta\! d\tau_2 \, I\Big((\bm x,t); (\bm x_1,t_1+i\tau_1);
(\bm x_2,t_2+i\tau_2)\Big)\nonumber\\
=-i\int_{-\infty}^{t_2}\! dt'' \bigg[I\Big((\bm x,t); (\bm
x_1,t_1+i\tau_1);
(\bm x_2,t''+i\tau_2)\Big)\bigg]\Bigg\vert_{\tau_2=0}^{\tau_2=\beta}\nonumber\\
=-i\int_{-\infty}^{t_2}\! dt'' 
\bigg[ \Big\langle \hat{X}(\bm x,t)\, \tilde{T}\, \hat{Y}(\bm x_1,t_1+i\tau_1)
\hat{Z}(\bm x_2,t''+i\beta)\Big\rangle_l  \nonumber\\
 -\Big\langle
\hat{X}(\bm x,t)\, \tilde{T}\, \hat{Y}(\bm x_1,t_1+i\tau_1)
\hat{Z}(\bm x_2,t'')\Big\rangle_l  \nonumber\\
-\big\langle \hat{Y}(\bm x_1,t_1)\big\rangle_l  
 \Big\langle \hat{X}(\bm x,t) \hat{Z}(\bm x_2,t''+i\beta)\Big\rangle_l
 +\big\langle \hat{Y}(\bm x_1,t_1)\big\rangle_l  
 \Big\langle \hat{X}(\bm x,t) \hat{Z}(\bm x_2,t'')\Big\rangle_l\nonumber\\
- \big\langle \hat{X}(\bm x,t)\big\rangle_l
 \Big\langle\tilde{T}\,\hat{Y}(\bm x_1,t_1+i\tau_1)\hat{Z}(\bm x_2,t''+i\beta) \Big\rangle_l \nonumber\\
+\big\langle \hat{X}(\bm x,t)\big\rangle_l
\Big\langle\tilde{T}\,\hat{Y}(\bm x_1,t_1
+i\tau_1)\hat{Z}(\bm x_2,t'') \Big\rangle_l\bigg],
\eea
where we substituted Eq.~\eqref{eq:integrand_3_point} and canceled
the terms $\propto \big\langle \hat{Z}(\bm x_2,t'')
\big\rangle_l$. Using again the assumption~\eqref{eq:corr_vanish3} we
further modify Eq.~\eqref{eq:3_point_corr2}
\bea\label{eq:itgr_mod1}
&& \int_0^\beta\! d\tau_2 \, I\Big((\bm x,t); (\bm x_1,t_1+i\tau_1); (\bm x_2,t_2+i\tau_2)\Big)\nonumber\\
&& 
=\int_{-\infty}^{t_1}\! dt' \frac{d}{dt'}\int_0^\beta\! d\tau_2 \,
I\Big((\bm x,t); (\bm x_1,t'+i\tau_1); (\bm x_2,t_2+i\tau_2)\Big)\nonumber\\
&& 
=-i\frac{d}{d\tau_1} \int_{-\infty}^{t_1}\! dt'\int_0^\beta\! d\tau_2 \, I\Big((\bm x,t); (\bm x_1,t'+i\tau_1); (\bm x_2,t_2+i\tau_2)\Big),
\eea 
therefore
\bea\label{eq:3_point_corr4}
\int_0^\beta\! d\tau_1 \int_0^\beta\! d\tau_2 \, I\Big((\bm x,t); (\bm x_1,t_1+i\tau_1); (\bm x_2,t_2+i\tau_2)\Big)\nonumber\\
=
-i\int_{-\infty}^{t_1}\! dt'\Bigg[\int_0^\beta\! d\tau_2 \, I\Big((\bm x,t); (\bm x_1,t'+i\tau_1); (\bm x_2,t_2+i\tau_2)\Big)\Bigg]\Bigg\vert_{\tau_1=0}^{\tau_1=\beta}\nonumber\\
=
-\int_{-\infty}^{t_1}\! dt'
\int_{-\infty}^{t_2}\! dt'' 
\bigg[ \Big\langle \hat{X}(\bm x,t)\, \tilde{T}\, \hat{Y}(\bm x_1,t'+i\tau_1)
\hat{Z}(\bm x_2,t''+i\beta)\Big\rangle_l  \nonumber\\
 -\Big\langle
\hat{X}(\bm x,t)\, \tilde{T}\, \hat{Y}(\bm x_1,t'+i\tau_1)\hat{Z}(\bm x_2,t'')\Big\rangle_l  \nonumber\\
- \big\langle \hat{X}(\bm x,t)\big\rangle_l \Big\langle\tilde{T}\,\hat{Y}(\bm x_1,t'+i\tau_1)\hat{Z}(\bm x_2,t''+i\beta) \Big\rangle_l
\nonumber\\
+\big\langle \hat{X}(\bm x,t)\big\rangle_l \Big\langle\tilde{T}\,\hat{Y}(\bm x_1,t'+i\tau_1)\hat{Z}(\bm x_2,t'') \Big\rangle_l\bigg]
\Bigg\vert_{\tau_1=0}^{\tau_1=\beta}\nonumber\\
=
-\int_{-\infty}^{t_1}\! dt'
\int_{-\infty}^{t_2}\! dt'' 
\bigg[ \Big\langle \hat{X}(\bm x,t)\, \tilde{T}\, \hat{Y}(\bm x_1,t'+i\beta)
\hat{Z}(\bm x_2,t''+i\beta)\Big\rangle_l  \nonumber\\
-\Big\langle \hat{X}(\bm x,t)\, \tilde{T}\, \hat{Y}(\bm x_1,t')
\hat{Z}(\bm x_2,t''+i\beta)\Big\rangle_l  \nonumber\\
-\Big\langle
\hat{X}(\bm x,t)\, \tilde{T}\, \hat{Y}(\bm x_1,t'+i\beta)\hat{Z}(\bm x_2,t'')\Big\rangle_l  \nonumber\\
 +\Big\langle
\hat{X}(\bm x,t)\, \tilde{T}\, \hat{Y}(\bm x_1,t')\hat{Z}(\bm x_2,t'')\Big\rangle_l  \nonumber\\
-\big\langle \hat{X}(\bm x,t)\big\rangle_l \Big\langle\tilde{T}\,\hat{Y}(\bm x_1,t'+i\beta)\hat{Z}(\bm x_2,t''+i\beta) \Big\rangle_l
\nonumber\\
+\big\langle \hat{X}(\bm x,t)\big\rangle_l \Big\langle\tilde{T}\,\hat{Y}(\bm x_1,t')\hat{Z}(\bm x_2,t''+i\beta) \Big\rangle_l
\nonumber\\
+\big\langle \hat{X}(\bm x,t)\big\rangle_l \Big\langle\tilde{T}\,\hat{Y}(\bm x_1,t'+i\beta)\hat{Z}(\bm x_2,t'') \Big\rangle_l
\nonumber\\
-\big\langle \hat{X}(\bm x,t)\big\rangle_l \Big\langle\tilde{T}\,\hat{Y}(\bm x_1,t')\hat{Z}(\bm x_2,t'') \Big\rangle_l\bigg],
\eea
where we substituted Eq.~\eqref{eq:3_point_corr2} and used the
relation~\eqref{eq:Y_l} to cancel the terms
$\propto \big\langle \hat{Y}(\bm x_1,t')\big\rangle_l $.  Next we
reorder the operators $\hat{Y}$ and $\hat{Z}$ in the
anti-chronological order and exploit the relation~\eqref{eq:KMS}
\bea\label{eq:3_point_corr5}
\int_0^\beta\! d\tau_1 \int_0^\beta\! d\tau_2 \, I
\Big((\bm x,t); (\bm x_1,t_1+i\tau_1); (\bm x_2,t_2+i\tau_2)\Big)\nonumber\\
=
-\int_{-\infty}^{t_1}\! dt'
\int_{-\infty}^{t_2}\! dt'' 
\bigg[ \frac{1}{2}\Big\langle \big\{\hat{Y}(\bm x_1,t'),\hat{Z}(\bm x_2,t'')\big\}\hat{X}(\bm x,t)\Big\rangle_l  \nonumber\\
-\Big\langle \hat{Z}(\bm x_2,t'')\, \hat{X}(\bm x,t)\, \hat{Y}(\bm x_1,t')\Big\rangle_l  \nonumber\\
-\Big\langle\hat{Y}(\bm x_1,t')\,
\hat{X}(\bm x,t)\,\hat{Z}(\bm x_2,t'')\Big\rangle_l  \nonumber\\
 +\frac{1}{2}\Big\langle
\hat{X}(\bm x,t)\, \big\{\hat{Y}(\bm x_1,t'),\hat{Z}(\bm
x_2,t'')\big\}\Big\rangle_l
\bigg]\nonumber\\
=
-\frac{1}{2}\int_{-\infty}^{t_1}\! dt'
\int_{-\infty}^{t_2}\! dt'' 
\bigg[ \Big\langle \hat{Y}(\bm x_1,t')\,\hat{Z}(\bm x_2,t'')\,\hat{X}(\bm x,t)\Big\rangle_l  \nonumber\\
+\Big\langle \hat{Z}(\bm x_2,t'')\, \hat{Y}(\bm x_1,t')\,\hat{X}(\bm x,t)\Big\rangle_l  \nonumber\\
-2\Big\langle \hat{Z}(\bm x_2,t'')\, \hat{X}(\bm x,t)\, \hat{Y}(\bm x_1,t')\Big\rangle_l  \nonumber\\
-2\Big\langle\hat{Y}(\bm x_1,t')\,
\hat{X}(\bm x,t)\,\hat{Z}(\bm x_2,t'')\Big\rangle_l  \nonumber\\
 +\Big\langle
\hat{X}(\bm x,t)\, \hat{Y}(\bm x_1,t')\,\hat{Z}(\bm x_2,t'')\Big\rangle_l  \nonumber\\
 +\Big\langle
\hat{X}(\bm x,t)\, \hat{Z}(\bm x_2,t'')\,\hat{Y}(\bm
x_1,t')\Big\rangle_l
\bigg].
\eea
The last expression can be cast in the form
\bea\label{eq:3_point_corr6}
\int_0^\beta\! d\tau_2 \int_0^\beta\! d\tau_2 \, I
\Big((\bm x,t); (\bm x_1,t_1+i\tau_1); (\bm x_2,t_2+i\tau_2)\Big)\nonumber\\
=
-\frac{1}{2}\int_{-\infty}^{t_1}\! dt'
\int_{-\infty}^{t_2}\! dt'' 
\bigg\{ \Big\langle \hat{Y}(\bm x_1,t')\big[\hat{Z}(\bm x_2,t''),\hat{X}(\bm x,t)]\big\rangle_l  \nonumber\\
+\Big\langle \hat{Z}(\bm x_2,t'')\big[ \hat{Y}(\bm x_1,t'),\hat{X}(\bm x,t)\big]\Big\rangle_l  \nonumber\\
 +\Big\langle
\big[\hat{X}(\bm x,t), \hat{Y}(\bm x_1,t')\big]\hat{Z}(\bm x_2,t'')\Big\rangle_l  \nonumber\\
 +\Big\langle
\big[\hat{X}(\bm x,t), \hat{Z}(\bm x_2,t'')\big]\hat{Y}(\bm x_1,t')\Big\rangle_l\bigg\} \nonumber\\
=
-\frac{1}{2}\int_{-\infty}^{t_1}\! dt'
\int_{-\infty}^{t_2}\! dt'' 
\bigg\{\Big\langle\big[
\big[\hat{X}(\bm x,t), \hat{Y}(\bm x_1,t')\big],\hat{Z}(\bm x_2,t'')\big]\Big\rangle_l  \nonumber\\
 +\Big\langle\big[
\big[\hat{X}(\bm x,t), \hat{Z}(\bm x_2,t'')\big],\hat{Y}(\bm x_1,t')\big]\Big\rangle_l \bigg\}.
\eea
Now for the three-point correlator~\eqref{eq:3_point_corr1} we obtain
\bea\label{eq:3_point_corr7}
\Big(\hat{X}(\bm x,t),\hat{Y}(\bm x_1,t_1),\hat{Z}(\bm x_2,t_2)\Big)
 =
\frac{1}{2\beta^2}\! \int_{-\infty}^{t_1}\! dt' \int_{-\infty}^{t_2}\! dt'' \, G^R_{\hat{X}\hat{Y}\hat{Z}}(\bm x,\bm x_1,\bm x_2;\, t,t',t''),
\eea
where we took into account that $t'\leq t_1 \leq t$ and $t''\leq t_2 \leq t$ and defined the
three-point retarded Green's function by
\bea\label{eq:retarded_corr3_point}
G^R_{\hat{X}\hat{Y}\hat{Z}}(\bm x,\bm x_1,\bm x_2;\, t,t',t'')=
-\frac{1}{2}\theta(t-t')\theta(t-t'')
\nonumber\\
\times \bigg\{\Big\langle\big[
\big[\hat{X}(\bm x,t), \hat{Y}(\bm x_1,t')\big],\hat{Z}(\bm x_2,t'')\big]\Big\rangle_l 
 +\Big\langle\big[
\big[\hat{X}(\bm x,t), \hat{Z}(\bm x_2,t'')\big],\hat{Y}(\bm x_1,t')\big]\Big\rangle_l \bigg\}.
\eea

Now consider a generic second-order transport coefficient given by the integral
\bea\label{eq:J_XYZ}
J[\hat{X},\hat{Y},\hat{Z}](\omega_1,\omega_2) = \beta^2\!\!\int\! d^3x_1\!\int\! d^3x_2\!
\int_{-\infty}^t\!\! dt_1\, e^{i\omega_1(t-t_1)}
e^{\varepsilon(t_1-t)}\nonumber\\
\times\int_{-\infty}^t\!\! dt_2\, e^{i\omega_2(t-t_2)}
e^{\varepsilon(t_2-t)}\Big(\hat{X}(\bm x, t),\hat{Y}(\bm x_1, t_1),\hat{Z}(\bm x_2, t_2)\Big),
\eea
where we introduced again nonzero frequencies $\omega_{1,2}> 0$ which
will be pushed to zero at the end.
From Eq.~\eqref{eq:3_point_corr7} we have
\bea\label{eq:J_XYZ1}
J[\hat{X},\hat{Y},\hat{Z}](\omega_1,\omega_2) 
=\frac{1}{2}
\int_{-\infty}^t\!\! dt_1\, e^{i\omega_1(t-t_1)}
e^{\varepsilon(t_1-t)}
\int_{-\infty}^t\!\! dt_2\, e^{i\omega_2(t-t_2)}e^{\varepsilon(t_2-t)}\nonumber\\
\times \int_{-\infty}^{t_1}\! dt' \int_{-\infty}^{t_2}\! dt''  \int\! d^3x_1\!\int\! d^3x_2\, G^R_{\hat{X}\hat{Y}\hat{Z}}(\bm x,\bm x_1,\bm x_2;\, t,t',t'').
\eea

In the uniform medium Green's function depends only on two space-time
arguments ($\bm x-\bm x_1$, $t-t'$) and ($\bm x-\bm x_2$, $t-t''$),
therefore we can define the Fourier transformation
\bea\label{eq:green_fourier_3point}
G^R_{\hat{X}\hat{Y}\hat{Z}}(\bm x, \bm x_1,\bm x_2;\, t, t', t'') =
\int\! \frac{d^3k_1}{(2\pi)^3}
\int\! \frac{d^3k_2}{(2\pi)^3}
\int_{-\infty}^{\infty}\frac{d\omega'}{2\pi}
\int_{-\infty}^{\infty}\frac{d\omega''}{2\pi}
\nonumber\\
e^{i[\omega' (t'-t)-\bm k_1\cdot (\bm x_1-\bm x)]}
e^{i[\omega'' (t''-t)-\bm k_2\cdot (\bm x_2-\bm x)]}
G^R_{\hat{X}\hat{Y}\hat{Z}}(\bm k_1,\bm k_2;\,\omega',\omega''),
\eea
which gives
\bea
&&\int\! d^3x_1\! \int\! d^3x_2\,  
G^R_{\hat{X}\hat{Y}\hat{Z}}(\bm x, \bm x_1,\bm x_2;\, t, t', t'')\nonumber\\ 
&& =
\int_{-\infty}^{\infty}\frac{d\omega'}{2\pi}
\int_{-\infty}^{\infty}\frac{d\omega''}{2\pi}
e^{i\omega' (t'-t)}e^{i\omega''(t''-t)}
G^R_{\hat{X}\hat{Y}\hat{Z}}(\omega',\omega''),
\eea
where
$G^R_{\hat{X}\hat{Y}\hat{Z}}(\omega',\omega'')\equiv\lim_{\bm k_{1,2}\to 0}
G^R_{\hat{X}\hat{Y}\hat{Z}}(\bm k_1,\bm k_2;\,\omega',\omega'')$.
Substituting this in
Eq.~\eqref{eq:J_XYZ1} we obtain
\bea\label{eq:J_XYZ2}
J[\hat{X},\hat{Y},\hat{Z}](\omega_1,\omega_2) 
= \frac{1}{2}
\int_{-\infty}^t\!\! dt_1\, 
e^{(\varepsilon-i\omega_1)(t_1-t)}
\int_{-\infty}^t\!\! dt_2\, 
e^{(\varepsilon-i\omega_2)(t_2-t)}\nonumber\\
\times 
\int_{-\infty}^{t_1-t}\! dt' 
\int_{-\infty}^{t_2-t}\! dt'' 
\int_{-\infty}^{\infty}\frac{d\omega'}{2\pi}
\int_{-\infty}^{\infty}\frac{d\omega''}{2\pi}
e^{i\omega' t'}e^{i\omega'' t''}
G^R_{\hat{X}\hat{Y}\hat{Z}}(\omega',\omega'')
\nonumber\\
=  \frac{1}{2}
\int_{-\infty}^{\infty}\frac{d\omega'}{2\pi}
\int_{-\infty}^{\infty}\frac{d\omega''}{2\pi} \,
G^R_{\hat{X}\hat{Y}\hat{Z}}(\omega',\omega'')
\nonumber\\
\times
\left[\int_{-\infty}^0\!\! dt_1\, 
e^{(\varepsilon-i\omega_1)t_1}
\int_{-\infty}^{t_1}\! dt' e^{i\omega' t'}\right]
\left[\int_{-\infty}^0\!\! dt_2\, 
e^{(\varepsilon-i\omega_2)t_2}
\int_{-\infty}^{t_2}\! dt'' e^{i\omega'' t''}\right],
\eea
where we performed subsequent variable changes $t'\to t'+t$,
$t''\to t''+t$ in the first step, and $t_1\to t_1+t$, $t_2\to t_2+t$
in the second step. The two inner integrals in the square brackets
should be computed according to Eq.~\eqref{eq:exp_integral}, thus
\bea\label{eq:J_XYZ3}
J[\hat{X},\hat{Y},\hat{Z}](\omega_1,\omega_2) 
= \frac{1}{2}\lim_{\delta' \to 0^+} \lim_{\delta'' \to 0^+}
\frac{i}{\omega_1+i\varepsilon}
\frac{i}{\omega_2+i\varepsilon}
\oint\frac{d\omega'}{2\pi i}
\oint\frac{d\omega''}{2\pi i} \,
G^R_{\hat{X}\hat{Y}\hat{Z}}(\omega',\omega'') \nonumber\\
\times \left(\frac{1}{\omega' -\omega_1-i(\varepsilon+ \delta')}
  -\frac{1}{\omega'-i \delta'}\right)
 \left(\frac{1}{\omega'' -\omega_2-i(\varepsilon+ \delta'')}-\frac{1}{\omega''-i \delta''}\right),
\eea
where both integrals are closed in the upper half-plane, where the
retarded Green's function is analytic.  Utilizing Cauchy's integral
formula and sequentially taking the limits
$\delta^{\prime \prime} \rightarrow 0^{+}, \delta^{\prime} \rightarrow
0^{+}$, and $\varepsilon \rightarrow$ $0^{+}$, we derive the following
result
\bea\label{eq:J_XYZ4}
J[\hat{X},\hat{Y},\hat{Z}](\omega_1,\omega_2) 
&=& \frac{1}{2} \lim_{\varepsilon \to 0^+}
\lim_{\delta' \to 0^+}
\frac{i}{\omega_1+i\varepsilon}
\frac{i}{\omega_2+i\varepsilon}\nonumber\\
&&\hspace{-2cm}\oint\frac{d\omega'}{2\pi i}
\left(\frac{1}{\omega' -\omega_1-i(\varepsilon+ \delta')}
-\frac{1}{\omega'-i \delta'}\right)
\Big[ G^R_{\hat{X}\hat{Y}\hat{Z}}(\omega',\omega_2+i\varepsilon)
-G^R_{\hat{X}\hat{Y}\hat{Z}}(\omega',0)\Big] \nonumber\\
&&= \frac{1}{2} \lim_{\varepsilon \to 0^+}
\frac{i}{\omega_1+i\varepsilon}
\frac{i}{\omega_2+i\varepsilon}\nonumber\\
&&\hspace{-2cm}
\Big[ G^R_{\hat{X}\hat{Y}\hat{Z}}(\omega_1+i\varepsilon,\omega_2
+i\varepsilon)-G^R_{\hat{X}\hat{Y}\hat{Z}}(\omega_1+i\varepsilon,0)
-G^R_{\hat{X}\hat{Y}\hat{Z}}(0,\omega_2+i\varepsilon)+G^R_{\hat{X}\hat{Y}\hat{Z}}(0,0)\Big]\nonumber\\
&&\hspace{-2cm}= \frac{1}{2}
\frac{i}{\omega_1}\frac{i}{\omega_2}
\Big[ G^R_{\hat{X}\hat{Y}\hat{Z}}(\omega_1,\omega_2)-G^R_{\hat{X}\hat{Y}\hat{Z}}(\omega_1,0)
-G^R_{\hat{X}\hat{Y}\hat{Z}}(0,\omega_2)+G^R_{\hat{X}\hat{Y}\hat{Z}}(0,0)\Big].
\eea
 Taking the limit $\omega_{1,2} \rightarrow 0$  (zero-frequency, we arrive at the final expression
\bea\label{eq:J_XYZ5}
J[\hat{X},\hat{Y},\hat{Z}](0,0)
=\beta^2\!\!\int\! d^4x_1 d^4x_2
\Big(\hat{X}(x),\hat{Y}(x_1),\hat{Z}(x_2)\Big)
=-\frac{1}{2}\frac{\partial}{\partial\omega_1}\frac{\partial}{\partial\omega_2}
G^R_{\hat{X}\hat{Y}\hat{Z}}(\omega_1,\omega_2)\bigg\vert_{\omega_{1,2}=0}
\eea
with 
\bea\label{eq:green_fourier3_point1}
G^R_{\hat{X}\hat{Y}\hat{Z}}(\omega_1,\omega_2) 
&=& \int_{-\infty}^{\infty}\!\! dt_1 e^{-i\omega_1 (t_1-t)}\!\!
\int_{-\infty}^{\infty}\!\! dt_2 e^{-i\omega_2 (t_2-t)}
\int\! d^3x_1 \int\! d^3x_2\,  
G^R_{\hat{X}\hat{Y}\hat{Z}}(\bm x, \bm x_1,\bm x_2;\, t, t_1, t_2)\nonumber\\
&=& -\frac{1}{2}
\int_{-\infty}^{0}\!\! dt_1 e^{-i\omega_1 t_1}\!\! \int_{-\infty}^{0}\!\! dt_2 e^{-i\omega_2 t_2}\!\! \int\! d^3x_1\! \int\! d^3x_2\, \nonumber\\ 
&& \hspace{-2.5cm}\times
\bigg\{
\Big\langle\big[\big[\hat{X}(\bm 0, 0),
\hat{Y}(\bm x_1,t_1)\big],\hat{Z}(\bm x_2,t_2)\big]\Big\rangle_l 
 +\Big\langle\big[\big[\hat{X}(\bm 0, 0),
\hat{Z}(\bm x_2,t_2)\big],\hat{Y}(\bm x_1,t_1)\big]\Big\rangle_l \bigg\},
\eea
where we substituted Eq.~\eqref{eq:retarded_corr3_point}, used the
homogeneity of the system and performed variable changes
$t_1\to t_1+t$, $t_2\to t_2+t$, $\bm x_1\to \bm x_1+\bm x$,
$\bm x_2\to \bm x_2+\bm x$ to obtain the last expression.
Note that in Eq.~\eqref{eq:J_XYZ5} the Green's
function should be evaluated in the fluid rest frame.  

From Eqs.~\eqref{eq:green_fourier3_point1} and \eqref{eq:J_XYZ4} we find
\bea\label{eq:parity_green3}
\big\{G^R_{\hat{X}\hat{Y}\hat{Z}}(\omega_1,\omega_2)\big\}^* &=&
G^R_{\hat{X}\hat{Y}\hat{Z}}(-\omega_1, -\omega_2),\\
\label{eq:parity_trans3}
\big\{J[\hat{X},\hat{Y},\hat{Z}](\omega_1,\omega_2)\big\}^* &=&
J[\hat{X},\hat{Y},\hat{Z}](-\omega_1,-\omega_2),
\eea
as for hermitian operators we have the property 
\bea
\Big\langle\big[\big[\hat{X}(\bm 0, 0),
\hat{Y}(\bm x_1,t_1)\big],\hat{Z}(\bm x_2,t_2)\big]\Big\rangle_l^*
=\Big\langle\big[\big[\hat{X}(\bm 0, 0),
\hat{Y}(\bm x_1,t_1)\big],\hat{Z}(\bm x_2,t_2)\big]\Big\rangle_l.
\eea
From Eq.~\eqref{eq:parity_green3} we have also
\bea\label{eq:parity_green1}
{\rm Re}G^R_{\hat{X}\hat{Y}\hat{Z}}(-\omega_1,-\omega_2) &=&
{\rm Re}G^R_{\hat{X}\hat{Y}\hat{Z}}(\omega_1,\omega_2),\\
{\rm Im}G^R_{\hat{X}\hat{Y}\hat{Z}}(-\omega_1,-\omega_2) &=&
-{\rm Im}G^R_{\hat{X}\hat{Y}\hat{Z}}(\omega_1,\omega_2),
\eea
therefore from Eq.~\eqref{eq:J_XYZ5} we obtain in
the zero-frequency limit
\bea\label{eq:J_XYZ6}
J[{\hat{X},\hat{Y},\hat{Z}}](0,0)
=-\frac{1}{2}\frac{\partial}{\partial\omega_1}\frac{\partial}{\partial\omega_2}{\rm
  Re}G^R_{\hat{X}\hat{Y}\hat{Z}}(\omega_1,\omega_2)\bigg\vert_{\omega_{1,2}=0}.
\eea


\end{document}